\documentclass[pre,aps,10pt,twocolumn,twoside,floats,preprintnumbers,floatfix,superscriptaddress,a4paper,showpacs,showkeys,longbibliography]%
{revtex4-1}

\usepackage{amsmath}
\usepackage{amsfonts}
\usepackage{amssymb}

\usepackage{hyperref}

\usepackage{times}
\usepackage{time}
\usepackage{graphicx}

\usepackage{color}

   \newcommand{\der}[2]{\frac{d{#1}}{d{#2}}}

\def\ie{\emph{i.e.}~}
\def\cf{\emph{cf.}~}
\def\eg{\emph{e.g.}~}

\def\fig#1{\ref{Fig:#1}}
\def\Fig#1{Fig.~\fig{#1}}

\def\eq#1{(\ref{Eq: #1})}
\def\Eq#1{Eq.~\eq{#1}}
\def\Eqs#1{Eqs.~\eq{#1}}

\def\Sect#1{Sect.~\ref{sec:#1}}

\def\rmd{\ensuremath{\mathrm{d}}}

\def\celsius{\ensuremath{^{\circ}\text{C}}}

\begin{document}
\title{Impact of microphysics on the growth of one-dimensional breath figures}

\author{L. Stricker}
\affiliation{\text{Max Planck Institute for Dynamics and Self-Organization (MPI DS), 37077 G\"ottingen, Germany}}

\author{J. Vollmer}
\affiliation{\text{Max Planck Institute for Dynamics and Self-Organization (MPI DS), 37077 G\"ottingen, Germany}}
\affiliation{\text{Faculty of Physics, Georg-August Univ.~G\"ottingen, 37077 G\"ottingen, Germany}}

\date{\today -- \now}


\begin{abstract}
  Droplet patterns condensing on solid substrates (breath figures)
  tend to evolve into a self-similar regime, characterized by a
  bimodal droplet size distribution. The distributions comprise a
  bell-shaped peak of monodisperse large droplets, and a broad range
  of smaller droplets. The size distribution of the latter follows a
  scaling law characterized by a non-trivial polydispersity
  exponent. We present here a numerical model for three-dimensional
  droplets on a one-dimensional substrate (fiber) that accounts for
  droplet nucleation, growth and merging.  The
  polydispersity exponent retrieved using this model is not universal.
  Rather it depends on the microscopic details of droplet nucleation and merging.
  In addition, its values consistently differ from the theoretical prediction by Blackman (Phys.  Rev. Lett.,
  2000). Possible causes of this discrepancy are pointed out.
\end{abstract}

\maketitle

\section{Introduction}
\label{sec:Introduction}

When a flux of supersaturated vapor gets into contact with a solid
substrate, a condensation process can originate, leading to the
formation of droplets patterns on the substrate (``breath
figures'' \cite{ray11}).
The interest for breath figures is both theoretical and practical.
From the theoretical point of view, they can be used as a test
ground for scaling concepts in a well-posed non-equilibrium
setting. From the practical point of view, they appear in many
natural phenomena, \eg dew deposition on a spider net or on a
leaf. They can also be exploited for technological applications:
water collection from dew harvesting \cite{nik96, clu08, lek11},
biological sterilization \cite{bey06}, manufacturing of surface
structures and patterns for nano-technologies \cite{bok04, hau04,
wan07, ryk11}, fabrication of efficient heat-exchangers and
cooling devices \cite{sik11, ros02, lea06}. For such applications,
understanding and controlling the droplets formation, growth and
coalescence is crucial.

The formation of breath figures undergoes several phases
\cite{bey91}. First, the droplets nucleate on the substrate; then
they grow and coalesce, creating a roughly monodisperse
distribution. Eventually the space released by merging is
sufficient for the nucleation of new droplets. As the evolution
continues, self-similar droplet patterns appear
\cite{vio88,fam88,kol89, bri98}.

In this phase, bimodal droplet size distributions emerge in
experiments \cite{bey86,vio88,car97,had98,bla12}, as well as in
simulations \cite{fam89,ulr04,bla12}. The size distributions
feature a monodisperse bell-shaped peak for the largest droplets,
and a power-law distribution for the smaller droplets,
characterized by a non-trivial polydispersity exponent. Scaling
descriptions for the droplet number density have been largely
adopted in the classical theory for breath figures
\cite{vio88,fam88,fam89,fam89b,mea89,mea91,mea92}. Such a theory
relates the polydispersity exponent to the exponents for the time
decay of the droplet number and the porosity, \ie the fraction of
the non-wetted area over the total area of the substrate
\cite{kol89,bri98}. The scaling descriptions of the droplet size
distribution are also solutions \cite{cue97,cue98} of the
Smoluchowski coagulation equation \cite{smo16}, an
integro-differential equation describing the evolution of the
droplet size distribution as a consequence of merging processes.
However, for droplets growing on two-dimensional substrates the
polydispersity exponent derived from the solution of the
Smoluchowski coagulation equation \cite{bla00} was found to be
noticeably smaller than the value found in numerical simulations
\cite{bla10,bla12} and larger than the one observed in experiments
\cite{bla12}.

In the present paper, we examine the case of a one-dimensional
substrate with three-dimensional droplets. We use a numerical
approach in order to test the existing scaling theory and, in
particular, the common assumption that the polydispersity exponent
takes a universal value
\cite{vio88,fam88,fam89,fam89b,mea89,mea91,mea92,bla00}. We
introduce a numerical model based on a nucleation rule governed by
surface tension-driven instabilities. The dynamics account for the
presence of a precursor film between droplets, droplet growth due
both to direct mass deposition from the surrounding vapor and
surface diffusion, and non-trivial droplet interactions due to
deviations from the spherical shape. By means of simulations, we
systematically explore the dependence of the droplet patterns on
the deposition rate, the rules of droplet interaction, the radius
of the fiber and the nucleation radius.
Surprisingly, we find a dependence of the polydispersity exponent
of the droplet size distribution on the microscopic details of the
model, leading to the conclusion that the polydispersity exponent
is not universal, as assumed in the classical scaling models.
Moreover, we observe a sizable mismatch between the predicted
\cite{bla00} and the observed values of polydispersity exponent.
We point out possible sources of the discrepancies.

The paper is organized as follows. In \Sect{Model} we introduce
our model and its numerical implementation.  In \Sect{Theory} we
revisit the scaling description of breath figures with special
emphasis on the predictions for droplet growth on fibers.  These
predictions are then, in \Sect{Results}, carefully tested by
comparison to comprehensive numerical data. Finally, we conclude
in \Sect{Conclusion}.

\section{Model and numerical method}
\label{sec:Model}

The precise dynamics of droplet nucleation and growth on a solid
substrate is still a debated matter and depends on the specific
system. Possible mechanisms include \cite{bey91,bey06}
heterogeneous nucleation on impurities and defects of the
substrate, homogeneous nucleation, hydrodynamic instabilities,
surface diffusion on the substrate, direct condensation of the
vapor molecules on the droplets surface, coalescence with
neighboring droplets, heat transfer localized at the triple line
and preventing nucleation. The radius of an isolated droplet on a
substrate increases following a power law in time, whose exponent
changes according to the specific growth mechanism
\cite{ros73,bey91,uca12}. However, in densely populated droplets
systems, the growth is always dominated by coalescence. Therefore,
the growth exponent of the average radius depends only on the
dimensionalities of the droplets and the substrate \cite{fam88}.

\subsection{Modeling breath figures on a fiber}

\begin{figure}
\includegraphics[width=0.37\textwidth]{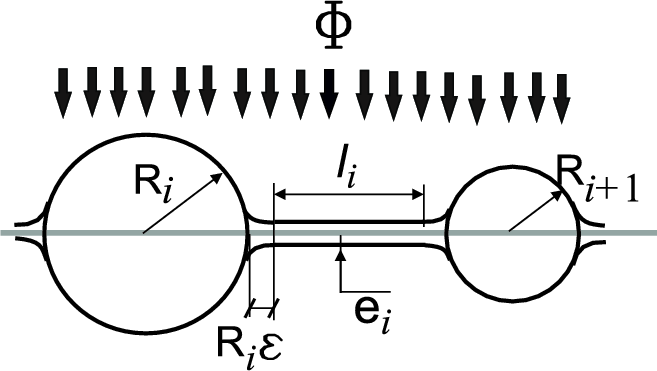}
\caption{Sketch of the model adopted for the numerical simulation
of the growth of droplets $i$ and $i+1$ with radii $R_i$ and
$R_{i+1}$, respectively. The droplets reside on a fiber (solid
gray line). They grow due to a constant water flux per unit
length, $\Phi$ [$\mu$m$^2$s$^{-1}$], due to the supersaturated
vapor impinging on the fiber. There is a precursor film of radial
thickness $e_i$ in the region between adjacent droplets, which has
a length $l_i$. When the precursor film collects sufficient water
volume, new droplets are nucleated. When two droplets approach
each other within their interaction ranges $\varepsilon R_i$ and
$\varepsilon R_{i+1}$, they merge.}
 \label{Fig:SketchDroplets1D}
\end{figure}

We model breath figures on a fiber as a one-dimensional chain of
spherical droplets, in contact with a supersaturated vapor. The
$i^{th}$ droplet has a radius $R_i$ and an interaction range
$\varepsilon R_i$, where $\varepsilon$ is constant for all the
droplets.  Such an interaction range has been introduced to keep
into account the deviation from the spherical shape observed in
experiments \cite{hal02} (see \Fig{SketchDroplets1D}). We account
for a cylindrical water prewetting film surrounding the fiber, in
the spaces between the droplets.  Water is deposited on the fiber
by condensation from the surrounding vapor. We do not consider
removal of water from the fiber by evaporation and gravity.

We consider a case where the supersaturation of the vapor is small
and the diffusive transport to the thread is the growth limiting
process for the droplets. In this case we expect cylindrically
symmetric vapor concentration profiles in the far field.
 We assume that the system is in thermal steady-state conditions, \emph{i.e.} the
water flux deposited on the fiber $\Phi$ -- water volume per unit
length per unit time -- is small enough to keep the temperature of
the droplets and the fiber uniform and invariant in time. We also
assume hydrodynamic steady-state conditions, taking $\Phi$ to be
constant in time and independent of the position, $x$, on the
fiber. These approximations have been largely adopted in the study
of breath figures \cite{vio88} and can be considered reasonable
for laminar vapor flows. The modelling of more complicated vapor
flows may require the inclusion of correction terms, but this is
beyond the scope of the present paper.

\subsection{Evolution of the droplets}

We adopt an event-driven approach and periodic boundary conditions
for the fiber. In order to simplify the problem, we decouple the
treatment of the droplets growth due to the impinging flux from
the nucleation process. In particular, for each time step, we
consider the growth in a time-continuous fashion and the
nucleation in a time-discrete fashion. The algorithm to advance
from time $t_j$ to the next instant $t_{j+1}$ proceeds as follows.
At first, we only consider the droplets growth due to the water
flux impinging on the droplets themselves and we disregard the
nucleation of new droplets. The constant flux density $\Phi$
impinging on a length $2R_i$ covered on the fiber by the particle
$i$ results in a growth law
\begin{equation}
  \der{}{t} \left( \frac{4}{3} \pi R_i^3 \right) = 2R_i\Phi
  \label{Eq: growth law of droplets}
\end{equation}
We integrate \Eq{growth law of droplets} in time, from $t$
to $t + \Delta t$,  finding
\begin{equation}
  R_i(t + \Delta t) = \sqrt{\frac{\Phi \Delta t}{\pi} + R_i^2(t)}
  \, .
 \label{Eq: R(t)}
\end{equation}
We calculate the time intervals $\Delta t_i$ for binary merging
events between adjacent droplets $i$ and $i+1$, by solving the
system
\begin{equation}
  x_{i+1} - x_i = (1+\varepsilon)
  \left[ R_{i+1}(t + \Delta t_i) + R_i(t + \Delta t_i) \right]
 \label{Eq: Find t intervals for collisions}
\end{equation}
where $x_i$ and $x_{i+1}$ are the positions of the respective
centers (note that they only move when the droplets merge). We
take the minimum of these time intervals, $\Delta
t_{\text{merge}}= \min{\Delta t_i}$, as a first estimate of the
time step to advance the system in time. This time step
corresponds to the first merging event that would take place if
there was no nucleation and no water deposited between the
droplets. We calculate the volume of water $V_{\text{gap},i}$
deposited on the gaps between adjacent droplets $i$ and $i+1$
during the time $\Delta t_{\text{merge}}$, by integrating
\begin{equation}
    \der{V_{\text{gap},i}}{t} = \Phi [x_{i+1} - x_i - R_{i+1} - R_i]
  \label{Eq: volume deposited between droplets}
\end{equation}
between $t$ and $t + \Delta t_{\text{merge}}$.

\begin{figure*}
\centering
\includegraphics[width=0.85\textwidth]{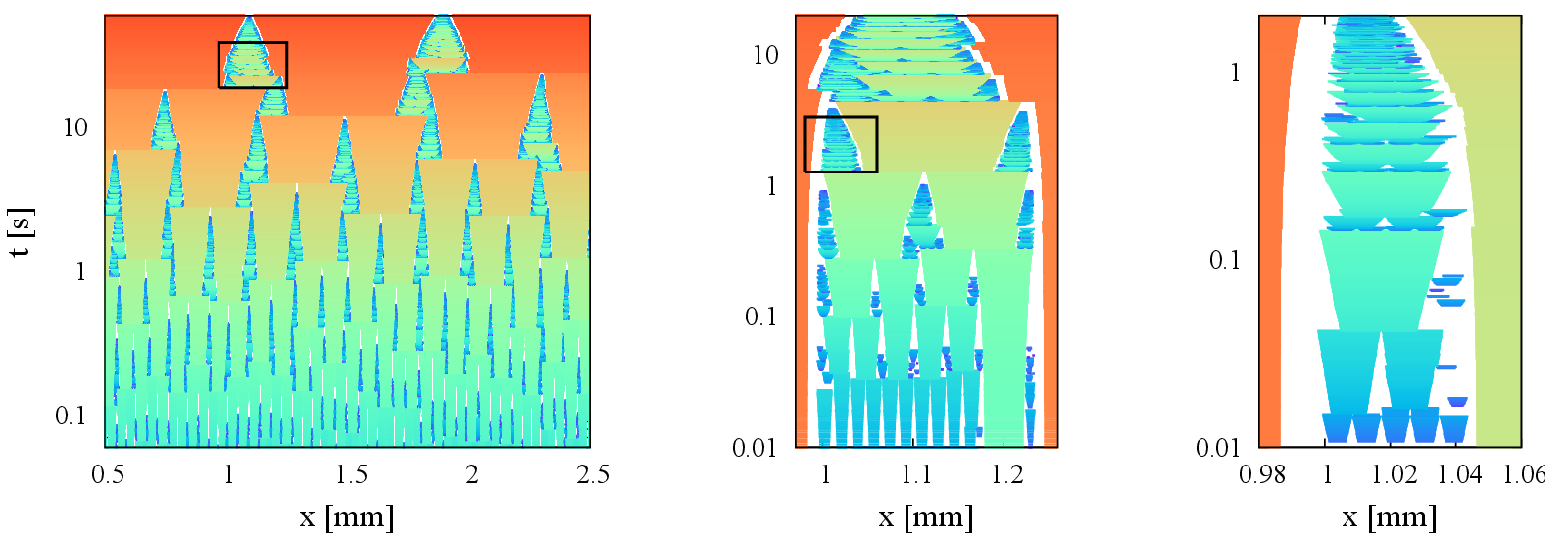}
\caption{(Color online.) Time evolution of the space occupied by
the droplets for
 $\varepsilon = 2\% $,
 $\Phi = 1000 \,\mu$m$^2 / $s,
 $N_0 = 1.5 \times 10^5$, and
 $R_f = R_{\text{min}} = 1\,\mu$m.
The different colors represent the different ages of the droplets
at the considered time instant and they vary from dark (blue
online), for the youngest and still very small droplets to
lit-gray (yellow, then red online), for the oldest hence largest
droplets. The central panel is a magnification of the area
indicated by the black frame in the left panel, and the right
panel shows a magnification of the frame in the central panel.
They are both displayed on a logarithmic scale with the origin
corresponding to the instant when the free area started to be
populated (\ie the bottom side of the black frame in the previous
panel). }
 \label{Fig:radii-time}
\end{figure*}

\subsection{Nucleation of droplets}

In each time step we determine if one or more nucleation events could happen
during the time interval $\Delta t_{\text{merge}}$. To this
aim, we consider the existence of a cylindrical precursor film,
surrounding the parts of the fiber where no droplets are present.
The precursor film grows in time by deposition of mass and may
eventually develop into a surface tension-driven instability.
Stability analysis \cite{gen04} for a fiber of radius $R_f$
reveals that perturbations with a wavelength $\lambda > 2\pi R_f$
are unstable.
The most unstable perturbation, \ie the fastest growing
perturbation, has a wavelength $\lambda^* = 2\sqrt{2}\pi R_f$
\cite{gen04}. Therefore, $\lambda^*$ can be regarded as a
characteristic wavelength of the system. Necklace-shaped chains of
droplets have been observed in experiments on fibers, where
equispaced droplets appeared at a distance $\lambda^*$ from each
other \cite{gen04}. In our model, we consider that a nucleation
event can take place between two droplets $i$ and $i+1$, during
the time $\Delta t_{\text{merge}}$ only if the gap length $l_i$
between such droplets, is larger than the characteristic
wavelength of the first unstable perturbation $\lambda^*$. The gap
length is defined as $l_i(t) = x_{i+1} - x_i - (1 +
\varepsilon)(R_{i+1} + R_i)$ and, in order to have a nucleation
event, it has to be
\begin{equation}
  l_i > \lambda^* = 2\sqrt{2}\pi R_f
  \label{Eq: cdt nucleation, enough space}
\end{equation}

From the length of the gap $l_i$, we determine the number of
possible equispaced nucleation sites in the gap, $N_i = \lfloor
l_i / \lambda^* \rfloor$. Additionally, we impose a minimum
droplet size $R_{\text{min}}$, for the nucleation event to take
place. We approximate the film between the droplets $i$ and $i +
1$ as a cylinder with volume
\begin{equation}
  V_{\text{film},i} = \pi \left[(R_f + e_i)^2 - R_f ^2 \right]
  l_i \, ,
  \label{Eq: volume of bridge, def}
\end{equation}
where $e_i$ is the thickness of the film. We check if the amount
of liquid $V_{\text{gap},i}$ deposited in the gap $l_i$ during
the time $\Delta t_{\text{merge}}$ is large enough to satisfy
the following condition:
\begin{equation}
  V_{\text{film},i} = V_{\text{film},i}^0 + V_{\text{gap},i}
  \geq \frac{4}{3}\pi R_{\text{min}}^3 N_i \, ,
  \label{Eq: cdt nucleation, enough vol}
\end{equation}
where $V_{\text{film},i}^0$ is the volume of the precursor film
in the $i^{th}$ gap at the beginning of the time step $\Delta
t_{\text{merge}}$, and $V_{\text{gap},i}$ is derived by
integrating \Eq{volume deposited between droplets} over $\Delta
t_{\text{merge}}$. In nucleation processes, the critical droplet
radius required to have a nucleation site can be calculated as
\cite{but03} $R_{\text{nucl}} = {2\sigma} / \left[{\rho_L R_V T
\textmd{ln}(p_V/p_\infty)}\right]$, in which $\sigma$ is the
surface tension between the liquid and the vapor, $\rho_L$ is the
density of the liquid, $T$ is the temperature expressed in Kelvin,
$R_V$ is the universal gas constant for water vapor,
$p_V/p_\infty$ is the supersaturation rate, with $p_V$ the
pressure of the vapor and $p_\infty$ the pressure of the saturated
vapor. In our model we adopt the minimum droplet size as
$R_{\text{min}} = R_{\text{nucl}}$. For typical environmental
temperatures ($20$\,--\,$40\celsius$) and supersaturation rates $p_V /
p_\infty \geq 1.001$ we have $R_{\text{nucl}} < 1\,\mu$m. If
\Eq{cdt nucleation, enough space} and \Eq{cdt nucleation, enough
vol} are both satisfied, a new nucleation event takes place. If
\Eq{cdt nucleation, enough space} is satisfied but \Eq{cdt
nucleation, enough vol} is not, we consider the volume of water
$V_{\text{gap},i}$ deposited on the $i^{th}$ gap as increasing the
thickness of the cylindrical precursor film, and we store the
information until the next time step.
 If neither \Eq{cdt nucleation, enough
space} nor \Eq{cdt nucleation, enough vol} are satisfied, we
consider the amount of water deposited on the gap $V_{\text{gap},i}$ as
collected by the droplets adjacent to the bridge itself. In
particular, this water volume will be collected by the largest
droplet, due to the Laplace pressure $p_L$ \cite{gen04}. Such a
pressure depends on the local curvature and, in the case of a
spherical droplet of radius $R$, it is given by $p_L = 2 \sigma /
R$.

\subsection{Merging of droplets}

We advance the system in time, by evolving it accordingly to the
estimated $\Delta t_{\text{merge}}$: we grow the droplets, we
insert the new nucleated ones and we merge the droplets when they
approach each other within their interaction ranges. In order to
do so, we replace the two merging droplets with a new one, with a
mass equal to the sum of the masses of the merging ones. Its
center is located in the center of mass of the two droplets. In
the simulation we explicitly prevent triple merging events between
neighboring droplets. Though rare, such unphysical events could
occur in the described numerical scheme, due to the collection of
liquid from the adjacent water bridges. When triple merging is at
hand at the end of the estimated time step $\Delta
t_{\text{merge}}$, we halve the estimated time step, and we evolve
the system accordingly.

\begin{figure}
\centering
\includegraphics[width=0.45\textwidth]{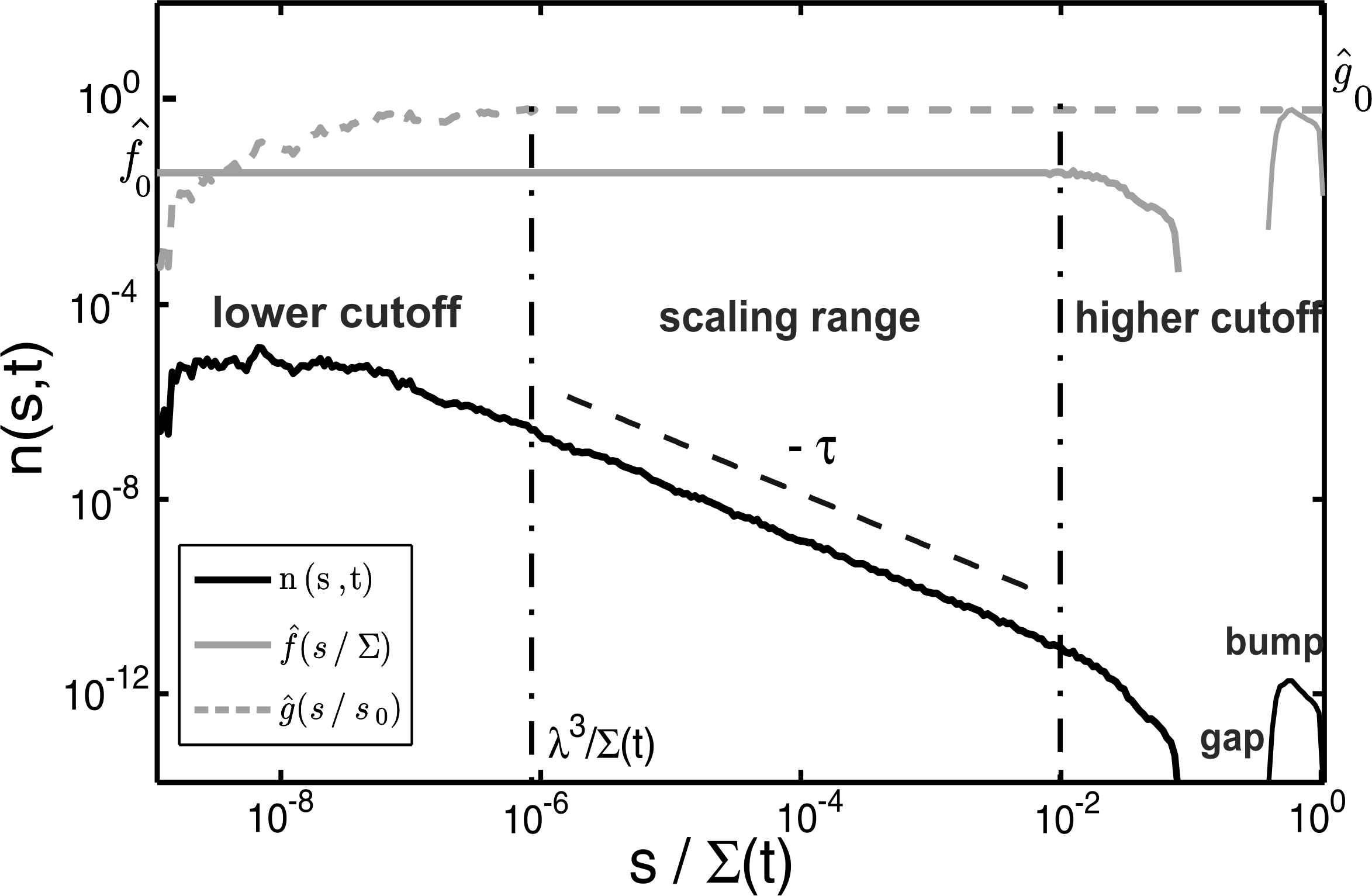}
\caption{Droplet number density for $\varepsilon = 2 \% $, $R_f =
R_{\text{min}} = 1 \,\mu$m, $\Phi = 1000 \,\mu$m$^2 / $s,
$L_f$ = 1 m, $N_0 = 1.5 \times 10^5$. The solid black line
represents $n(s,t)$. The dashed black line is a fitting line for
the polydisperse scaling range of the droplets and it has a slope
$-\tau$ on the log-log scale. The gray solid line represents the
cutoff for the large droplets $\hat{f}(s/ \Sigma(t))$. The gray
dashed line represents the cutoff for the small droplets:
$\hat{g}(s/ s_0)$. The two vertical dash-dotted lines delimit
the scaling range. For small droplets ($s<\lambda^*$) the
distribution is dominated by $\hat{g}(s/s_0)$. For large droplets
($s/\Sigma(t)>1\%$), the distribution is dominated by $\hat{f}(s/
\Sigma(t))$. In this range, we can identify a monodisperse
\emph{bump}, corresponding to the oldest droplets, and a
\emph{gap}, corresponding to the times where the spaces released
by merging droplets were not large enough to allow the nucleation
of new droplets. The present graphs are derived by averaging the
droplet size distributions over 10 time instants and over five
simulations with different but equivalent initial conditions. The
considered instants are in the self-similar regime ($t > 1000$ s)
and they are chosen in such a way to have 100 points per time
decade.}
 \label{Fig:n density and cutoffs}
\end{figure}
\section{Scaling theory for droplets on a fiber}
\label{sec:Theory}

In \Fig{radii-time} we show the time evolution of the droplets and
their coverage of the fiber surface. The horizontal colored
segments represent the areas covered by the droplets on the fiber
at a certain time $t$. Different colors reflect the different ages
of the droplets at the specific time $t$. In black (blue online)
we represent newly created small droplets, and the color fades
(turning to yellow, then red online) as the droplets become older.
When two droplets merge, they release two regions of length $\sim
R_1 + R_2 - \left( R_1^3 +
  R_2^3\right)^{1/3}$. For sufficiently big droplets the released space is
large enough to host new droplets. The opening gaps become larger
with the increase of the largest droplets. Eventually, new
droplets nucleate and grow in the gaps, forming a hierarchical
structure in the droplet size distribution. After a sufficiently
long time, self-similar droplet patterns emerge: pictures taken at
different time instants look alike, apart from length rescaling
(\cf\Fig{radii-time}). This is reflected in a power-law size
distribution of the middle-sized droplets
\cite{fam88, fam89, fam89b, mea89, mea91, mea92}, characterized by
an exponent $\tau$ (see \Fig{n density and cutoffs}), namely the
polydispersity exponent. On the other hand, the largest (oldest)
and the smallest (newest) droplets have a different behavior.
Their physics is captured in cutoff functions which describe the
termination of the power law for large and small droplets.
\cite{bla12}.
The small-scale cutoff accounts for the minimum droplet size $R_{\text{min}}$ and the
characteristic length of the surface tension-driven instability in the
precursor film $\lambda^*$.
The large-scale cutoff comprises a \emph{bump} and a \emph{gap}
(see \Fig{n density and cutoffs}). The bump represents the oldest
droplets in the system (first droplet population). The gap
originates from the times where the openings generated between the
first droplets were still too small to admit a second wave of
nucleation.

\subsection{Size of the largest droplets}

\begin{figure}
\includegraphics[width=0.4\textwidth]{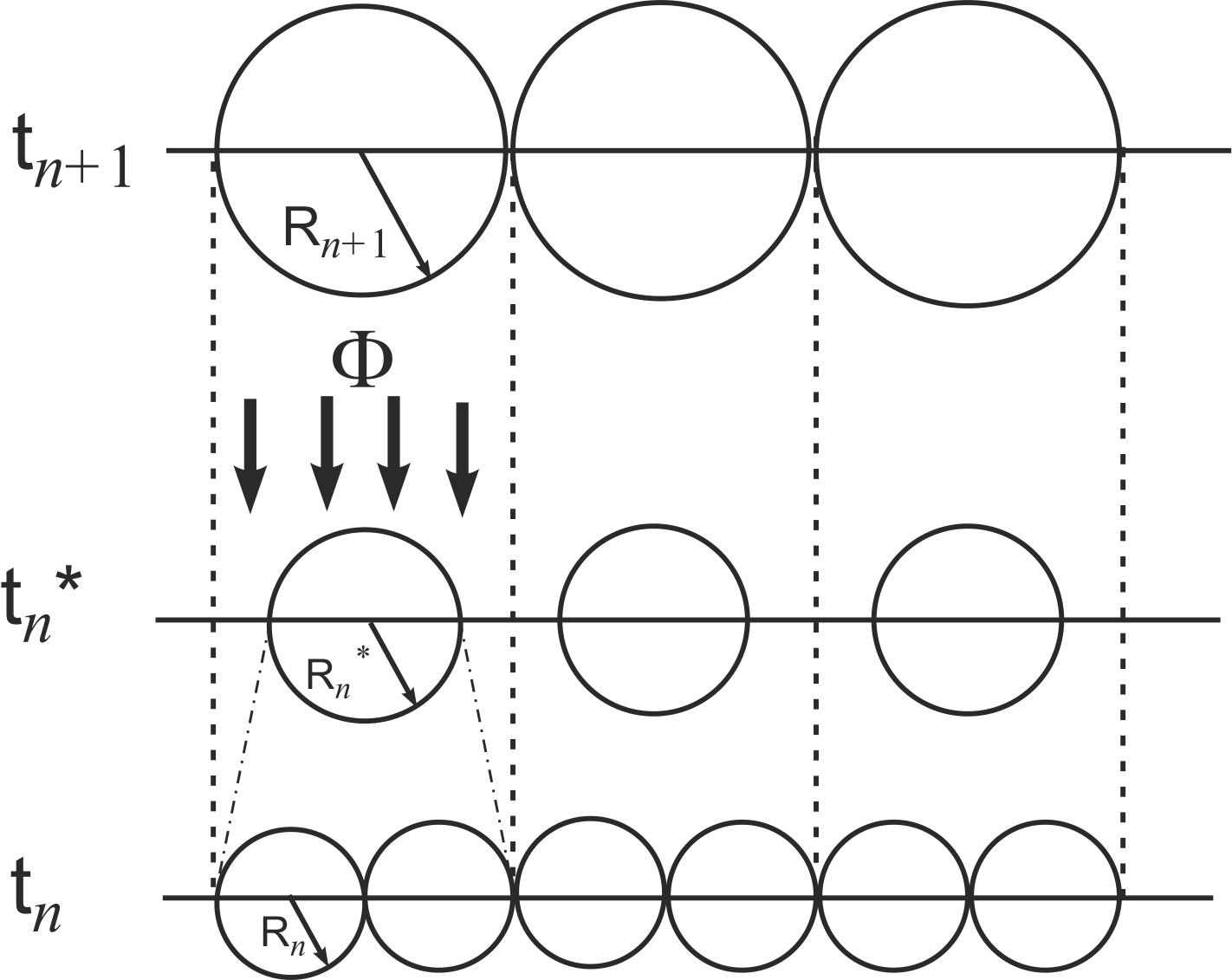}
\caption{Sketch of the simplified model adopted for deriving the
time growth law of the size of the largest droplet $\Sigma(t)$.
The exponent of such a time growth is calculated based on the
assumptions that the droplets are monodisperse and in touch with
each other. The considered mechanisms governing the growth are the
droplet coalescence two by two and the deposition of water from a
uniform external flux per unit length~$\Phi$.}
 \label{Fig:Sketch Rmax}
\end{figure}

We consider the following simplified setting, in order to give an
estimate of the growth law in time of the largest droplet size
$\Sigma(t)$. At time $t_n$, we take a chain of monodisperse
droplets of radius $R_n$ and size $\Sigma(t_n) = R_n^3$, in
contact with each other (see \Fig{Sketch Rmax}). At time $t_n^*$,
the droplets have merged two by two, therefore $\Sigma(t_n^*) =
2\Sigma(t_n)$. Since the merging process is almost instantaneous,
$t_n^* \simeq t_n$. After merging, the droplets start to grow as
an effect of the deposition of a uniform water flux per unit
length, $\Phi$, on the fiber. Due to mass conservation, at times
$t_n$ and $t_{n+1}$, it must be $2 R \Phi t = \frac{4}{3} \pi
R^3$. Hence,
\begin{equation} \Sigma = \left( \frac{3 \Phi} {2 \pi}
\right)^{3/2} t^z   \ \ \mbox{ with } z = 3/2 \, . \label{Eq:
t-Sigma}
\end{equation}
The so-calculated exponent $z$ is in agreement with both
experimental findings \cite{ste90} and theoretical predictions
\cite{fam88}. The prefactor $\left(\frac{3 \Phi} {2 \pi}
\right)^{3/2}$ has to be regarded as a lower bound. Since the
merging is assumed to happen instantaneously, the largest droplet
in the system at time $t_n^* \simeq t_n $ will most likely be a
droplet that has just originated from a merging. Hence, the upper
bound for the prefactor will be $ 2 \left(\frac{3 \Phi} {2 \pi}
\right)^{3/2}$.

\subsection{The droplet number density $n(s,t)$}

The droplet number density $n(s,t)$ is defined as the number of
droplets of size $s$ at time $t$ per size interval $\text{d}s$
of droplets and unit length of the substrate, where $s=r^3$ is the
size of a droplet with radius $r$. In our case, $n(s,t)$ has the
units of $[m]^{-4}$.
Consequently, the Buckingham-Pi theorem \cite{bar03} ensures that
the droplet number density can be expressed as $n(s,t) =
s^{-\theta} f(s/\Sigma(t), s/s_0)$, where $f(x)$ is a
dimensionless scaling function, $s_0$ characterizes the cutoff for
the small droplets, $\Sigma(t)$ is the maximum droplet size at
time $t$, and $\theta = (D+d)/D$ \cite{fam89} is an exponent
depending on the dimensionality of the droplets $D$ and of the
substrate $d$. Here, we consider the case of three-dimensional
droplets, $D=3$, growing on a one-dimensional fiber, $d=1$.
Therefore, from merely dimensional considerations one infers that
$\theta = 4/3$.

The classical scaling theory for breath figures \cite{fam88,
fam89} established that, in the late regime, the droplet size
distribution becomes self-similar: there is an increasing scaling
range between $s_0$ and $\Sigma(t)$, characterized by a
polydispersity exponent $\tau$ (\Fig{n density and cutoffs}). This
suggests that the function $f(s/\Sigma(t), s/s_0)$ can be
factorized into a power law \(
  \left({s}/{\Sigma(t)}\right)^{\theta - \tau}
\) and two cutoff functions, \( \hat{f}
\left({s}/{\Sigma(t)}\right) \) and \( \hat{g}
\left({s}/{s_0}\right) \), accounting for the large and the small
scales, respectively. For their asymptotics we request that $\hat
f(x) = \hat f_0 =$const for $x \ll 1$, and that $\hat g(x) = 1$
for $x \gg 1$.

Thus, the  droplet  number density  is expressed as
\begin{equation}
  n(s,t)
  = s^{-\theta}
  \left(\frac{s}{\Sigma(t)}\right)^{\theta - \tau}  \hat{f}
  \left(\frac{s}{\Sigma(t)}\right) \hat{g}
  \left(\frac{s}{s_0}\right) \, .
  \label{Eq: n(t,s)}
\end{equation}

In the framework of the renormalization group theory
\cite{bar96,gol92}, one would expect $\tau$ to be a universal
constant, depending only on the dimensionality of the system and
not on microscopic details \cite{vio88,fam88,fam89,fam89b,kol89,
mea89,mea91,mea92,cue97,bla00}. The value of $\tau$ is related to
the decaying exponents characterizing the time evolution of the
porosity and the number of droplets, by consistency reasons (see
Sections~\ref{sec:Ndroplets} and \ref{sec:Porosity}). This poses
some limitations to the range of the physically acceptable values
of $\tau$. Within such range, a theoretical derivation of $\tau$
has been proposed by Blackman and Brochard \cite{bla00}. In
Section~\ref{sec:Theoretical tau} we will revisit it, then we will
compare the expected value of $\tau$ to our numerical results
(Section~\ref{sec:Results}).

\subsection{The number of droplets per unit length, $N(t)$}
\label{sec:Ndroplets}

The total number of droplets $N(t)$ at time $t$ per unit length of
fiber can be written as
\begin{equation}
  N(t) = \int_0^\infty n(s,t) \: \mathrm{d}s
  \label{Eq: N(t), integral}
\end{equation}
To evaluate the integral we substitute \Eq{n(t,s)} into \Eq{N(t),
integral}, and we define a new variable $x \equiv s/\Sigma(t)$. We
replace the function inside the integral with a proper combination
of step functions, using the fact that $\hat{f}(x)$ should
contribute to the shape of the droplet distribution only for large
droplets, being a constant $\hat{f}_0$ otherwise. Similarly, the
role of $\hat{g}(x)$ is to provide a cutoff for small droplet
sizes. Therefore, it takes the value $ \hat{g} (x) = 1$ for all but
the smallest values of $x$. Consequently,
\begin{eqnarray}
  N(t)
  &=& \hat{f}_{0} \; \Sigma^{1-\theta} \int_{s_{N}/\Sigma(t)}^{x_{N}} x^{-\tau} \mathrm{d}x
  \nonumber \\[2mm]
  &=& \frac{\hat{f}_{0}}{1-\tau} \; \Sigma^{1-\theta} \;
  \left[ x_N^{1-\tau} - \left( \frac{s_N}{\Sigma(t)} \right)^{1-\tau} \right] \, ,
\label{Eq: N(t), integral 2}
\end{eqnarray}
where $x_N \simeq 1$ and $s_N \simeq s_0$ are constant.
Substituting \Eq{t-Sigma} into \Eq{N(t), integral 2} we find
\begin{equation}
  N(t)
  \simeq \frac{\hat{f}_0 }{1 - \tau}
  \left[ x_{N}^{1 - \tau} - \frac{s_{N}^{1 - \tau}}{\left(\frac{3\Phi}{2\pi} t \right)^{z(1 - \tau)}} \right]
  \left(\frac{3\Phi t }{2\pi} \right)^{z(1-\theta)} \, ,
\label{Eq: N(t), integral 3}
\end{equation}
and
for $t\rightarrow \infty$ the number of droplets hence decays in
time as
\begin{subequations}
\begin{equation}
  N(t) \sim t^{-z'}  \, ,
\label{Eq: N(t), exponential law}
\end{equation} with an exponent
\begin{equation}
  z' =
  \begin{cases}
    z(\theta - 1   ), & \mbox{if } \tau \leq 1  \, , \\
    z(\theta - \tau), & \mbox{if } \tau >    1  \, .
  \end{cases}
\label{Eq: exponent z' for time decay of N(t)}
\end{equation}

\label{Eq: N(t), total}
\end{subequations}

We note that the values of the exponent $z'$ do not depend on the
specific choice of $s_{N}$, provided that there is a sufficient
scale separation between $x_N$ and $s_N/\Sigma(t)$. The case of
$\tau \leq 1$ corresponds to a monodisperse droplet population.
The corresponding trivial exponent $z'= 1/2$ can also be found
from the consideration that an ideally monodisperse population of
droplets of size $\Sigma$ will cover the whole length of the fiber
$L_f$, in the limit $t\rightarrow \infty$. Therefore, in such a
case, one would have $1 \sim N \Sigma^{1/3} \sim N t^{1/2}$, such
that $N \sim t^{-1/2}$. In particular, this exponent describes the
decay of the number of large droplets populating the bump of the
roughly monodisperse large droplets in the large-scale cutoff
function (\Fig{n density and cutoffs}). For a polydisperse droplet
population $\tau > 1$, and \Eq{exponent z' for time decay of N(t)}
sets an upper limit for the range of the physically acceptable
values of $\tau$. The size of the smallest droplets is fixed and
the larger droplets grow. Hence, the total number of droplets must
decay and $z'>0$ (see \Eqs{N(t), total}). Since $z>0$, it must be
$1 < \tau < \theta$.

\subsection{The porosity $p(t)$}
\label{sec:Porosity}

The porosity $p$ is defined as $p = 1 - A_d / A_{\text{tot}}$, where
$A_d$ is the wetted area covered by the droplets and $A_{\text{tot}}$ is
the total area of the substrate. For one-dimensional fibers these
quantities correspond to the wetted length and the total length of
the fiber, respectively. It is convenient to define an effective
porosity $p^*$ that keeps into account the interaction ranges of
the droplets
\begin{equation}
  p^* = 1 - \sum\limits_{i=1}^N A_i (1+\varepsilon) / A_{\text{tot}} \, .
  \label{Eq: modified porosity def}
\end{equation}
Here, $A_i(1+\varepsilon)$ is the effective area occupied by the
$i^{th}$ droplet, \ie for a one-dimensional substrate, $A_i =
2R_i$, with $R_i$ the radius and $(\varepsilon R_i)$ the
interaction range of the $i^{th}$ droplet. In the general case
$A_i \sim s_i^{d/D}$ such that the effective porosity is
\begin{equation}
  p^*(t) \sim  1 - \int_0^\infty n(s,t) \; C s^{d/D} \; (1+\varepsilon) \: \mathrm{d}s \,
  ,
\label{Eq: effective p(t), integral}
\end{equation}
where $C$ is a constant depending on the geometry of the system.
Following a procedure similar to the one adopted to calculate the
number of droplets per unit length $N(t)$, and introducing the
exponent $\theta = (d+D) /D$, we find
\begin{eqnarray}
  p^*(t)
  = 1 - \frac{\hat{f}_{0}}{\theta-\tau} \; C\; (1+\varepsilon)
  \int_{s_{p}/\Sigma(t)}^{x_{p}} x^{\theta -\tau-1}
  \: \mathrm{d}x \, ,
  \label{Eq: effective p(t), integral 2}
\end{eqnarray}
where again $x_p \simeq 1$ and $s_p \simeq s_0$, even though they
may slightly differ from the integration limits used to calculate
the total number of droplets, $x_N$ and $s_N$. After few algebraic
steps, we derive
\begin{equation}
  p^*(t)
  = 1 - \frac{\hat{f}_0}{\theta - \tau}
  \left[ x_{{p}}^{\theta - \tau}
    - \frac{s_{{p}}^{\theta - \tau}}{\left(\frac{3\Phi}{2\pi} t \right)^{z(\theta - \tau)}} \right]
  C \; (1 + \varepsilon) \, .
\label{Eq: effective p(t), integral 3}
\end{equation}
In the late regime, $t \rightarrow \infty$, we expect $p^*(t)
\simeq 0$, because the number of gaps decays together with the
droplet number, \Eq{exponent z' for time decay of N(t)}, and the
length of the gaps is bounded in our system. Hence,
$1 - \hat{f}_{0} \; C \; (1+\varepsilon) \: x_{{p}}^{\theta - \tau} /(\theta -
\tau) = 0$ and
\begin{equation}
  p^*(t)
  \simeq
  \frac{\hat{f}_0 \; s_{{p}}^{\theta - \tau}}{(\theta - \tau) \left(\frac{3\Phi}{2\pi} t \right)^{z(\theta - \tau)}}
   \; C \; (1 + \varepsilon) \, .
\label{Eq: effective p(t), integral 4}
\end{equation}
Therefore, in the late-time scaling regime, the effective porosity
decays in time as
\begin{equation}
  p^*(t) \sim t^{-k}   \ \ \mbox{ with } k = z(\theta - \tau) \, .
  \label{Eq: exponent k, for time decay of effective porosity}
\end{equation}
Note that this exponent takes the same value as the exponent $z'$
determined for the $N(t)$, when $\tau \geq 1$ (\cf \Eqs{N(t),
total}). This indicates that the porosity can be viewed as the
number of gaps (coinciding with the number of droplets for a
one-dimensional system) multiplied by a characteristic gap size,
which takes a constant value in the long-time limit.

Interestingly, this result also holds when considering only
droplets larger than a fixed finite size $s_c > s_0$, \ie smaller
droplets are considered part of the larger gaps.
One can check by straightforward calculations that only the
prefactors of the asymptotic power laws are affected by this
change of the lower cutoff.

\subsection{Expected value of $\tau$}
\label{sec:Theoretical tau}

The value of $\tau$ has been theoretically determined by Blackman
and Brochard \cite{bla00} from an analysis of the scaling of
Smoluchovski's equation \cite{smo16} for droplet coagulation,
which is then evaluated in a renormalization group framework under
the assumption that the particle number and the porosity must obey
the same time dependence.

The derivation of Blackman and Brochard \cite{bla00} relies on the
assumption that the collision rate
$\dot{n}_{\text{coll}}(s_1,s_2;t)$, \ie the number density of
droplets of size $s_1$ and $s_2$ that collide at time $t$ per unit
length per unit time, factorizes into the product of single
particle distribution functions and a geometrical factor:
\begin{equation}
  \dot{n}_{\text{coll}}(s_1,s_2;t)
  \sim n(s_1,t) \frac{n(s_2,t)}{N(t)}
  \left(s_1^{-1/D} + s_2^{-1/D}\right)
  \label{Eq: Probability of collision, factorization}
\end{equation}
where $n_{\text{coll}}(s_1,s_2;t)$ is the number density of
collisions per unit length of substrate between two droplets of
size $s_1$ and $s_2$, from time $0$ to time $t$; $n(s_1,t)$ is the
number density of droplets of size $s_1$ per unit length of
substrate at time $t$ and it represents the probability density of
having a droplet of size $s_1$ at time $t$ (the normalization
constant $L_f$ can be seen as equivalent to the number of knots in
a Boltzmann lattice); $n(s_2,t) / N(t)$ is the probability density
of having a neighboring droplet of size $s_2$, uncorrelated to the
size $s_1$. The term $(s_1^{-1/D} + s_2^{-1/D})$ represents the
speed at which the edges of the two adjacent droplets approach
each other, by effect of the mass deposition through $\Phi$. In
order to demonstrate this, in the case of $D = 3$ and $d = 1$, one
can write such speed as $-\dot{l_1}$, where the gap between the
two adjacent droplets is $l_1 = x_2 - x_1 - (R_2 + R_1)(1 +
\varepsilon)$. By substituting \Eq{R(t)} and taking the derivative
with respect to time, one then finds $\dot{R}_i \sim s_i^{-D}$.

The total collision rate per unit length of substrate at time~$t$,
$\dot{N}_{\text{coll}}(t) = \int_0^\infty
\dot{n}_{\text{coll}}(s_i,s_j;t) \:\rmd s_i \: \rmd s_j$ is
expected to scale as~\cite{bla00} $\dot{N}_{\text{coll}}(t) \sim
N(t)$. By equating such a scaling assumption and \Eq{Probability
of collision, factorization}, upon substitution of \Eq{N(t),
total}, a relationship is found, among $\tau$, $d$ and $D$. A
similar procedure is repeated, keeping into account
$A_{\text{coll}}(t)$, the change in the substrate area covered by
the droplets per unit length, due to collision events, instead of
$N_{\text{coll}}(t)$. The combination of the relationships among
$\tau$, $d$ and $D$ obtained through such a scaling analysis
yields \cite{bla00} that $\tau_{\text{theor}} = 7/6$ for the
growth of three-dimensional droplets on a fiber. We refer the
reader to \cite{bla00} for the details of the derivation.

\section{Results}
\label{sec:Results}

\begin{figure}
\centering
\includegraphics[width=0.45\textwidth]{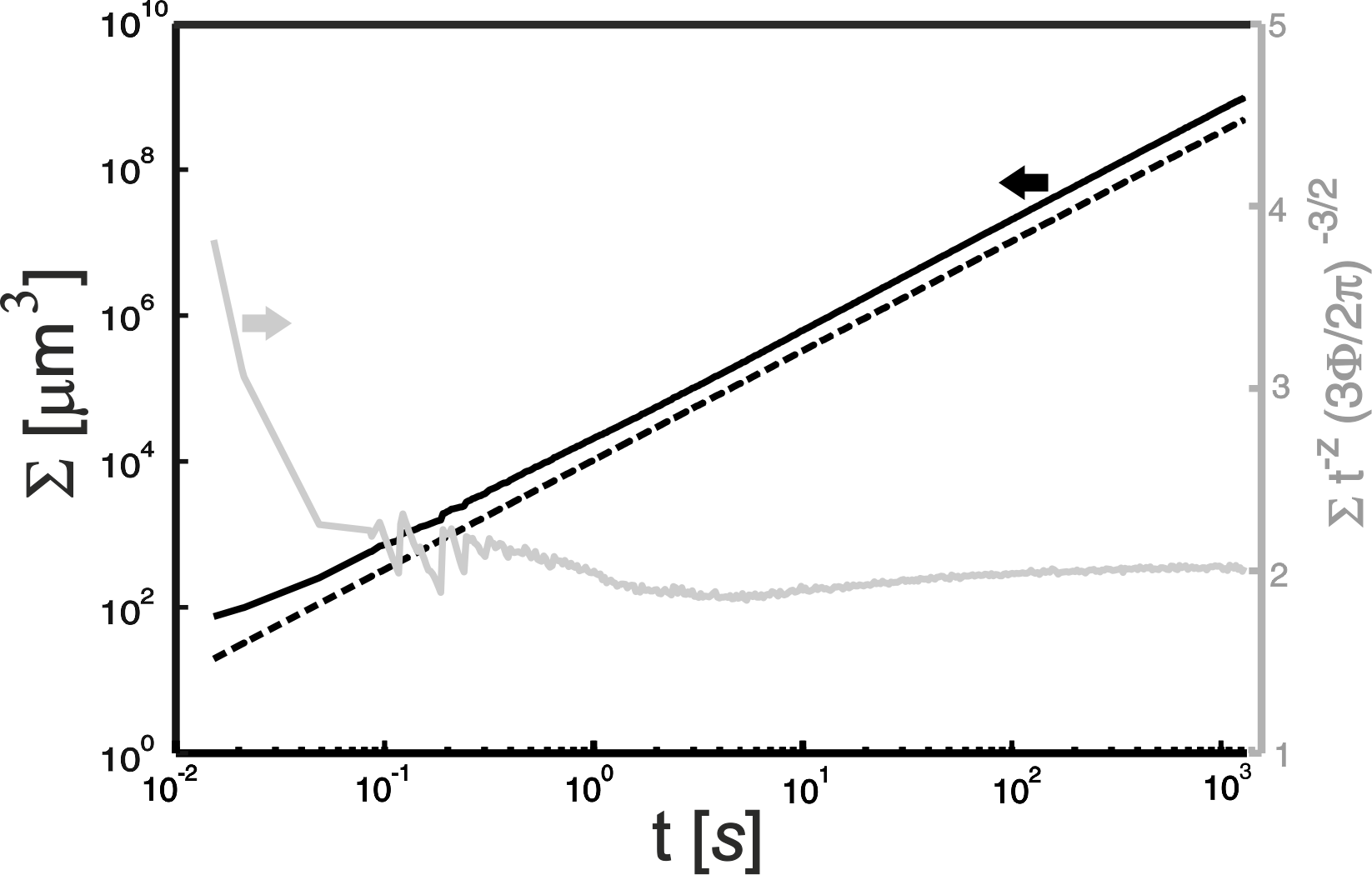}
\caption{Time evolution of the maximum droplet size
 (same parameters as in \Fig{n density and cutoffs}).
The black solid line represents the values inferred from our
simulations. The black dashed line has a slope of $3/2$ on the
double logarithmic axes, proving that $\Sigma \sim t^{3/2}$, and
it represents the function $y = (3 \Phi t / 2\pi)^z$. The gray
solid line is a reduced plot, obtained by dividing $\Sigma$ by the
RHS (right hand side) of \Eq{t-Sigma}.}
 \label{Fig:t-Sigma}
\end{figure}

\subsection{Setup of simulations and growth of the largest droplets}

All simulations start from an initial condition where $N_0 = 1.5
\times 10^5$ droplets form an equispaced necklace at the distance
$\lambda^*$ from each other; the length of the fiber $L_f$ is
adapted accordingly. The initial conditions differ by a slight
polydispersity of the droplet radii, which are chosen as $r_{0,i}
= R_{\text{min}} (1 + 0.01 \, I_{\text{rand},i} )$
where  $1\leq i \leq N_0 $, and $I_{\text{rand},i}$ are random
numbers in the interval $0 \le I_{\text{rand},i} \le 1$ .
We select the characteristic size of the distribution,
$\Sigma(t)$, to be the maximum droplet size at time $t$. As
predicted by \Eq{t-Sigma}, it scales as $\Sigma \sim
\left(\frac{3\Phi}{2\pi} t \right)^z$ , with $z = 3/2$
(\Fig{t-Sigma}). The asymptotic value of 2 in the reduced plot
indicates that most likely the largest droplet in the system has
recently undergone a collision.

\subsection{Scaling of the droplet number and the porosity}

We check the consistency of the exponents for
$p^*(t)$ and $N(t)$ by inspecting
\Eq{exponent k, for time decay of effective porosity} and
\Eq{exponent z' for time decay of N(t)}, respectively.

\begin{figure}
\centering
\includegraphics[width=0.45\textwidth]{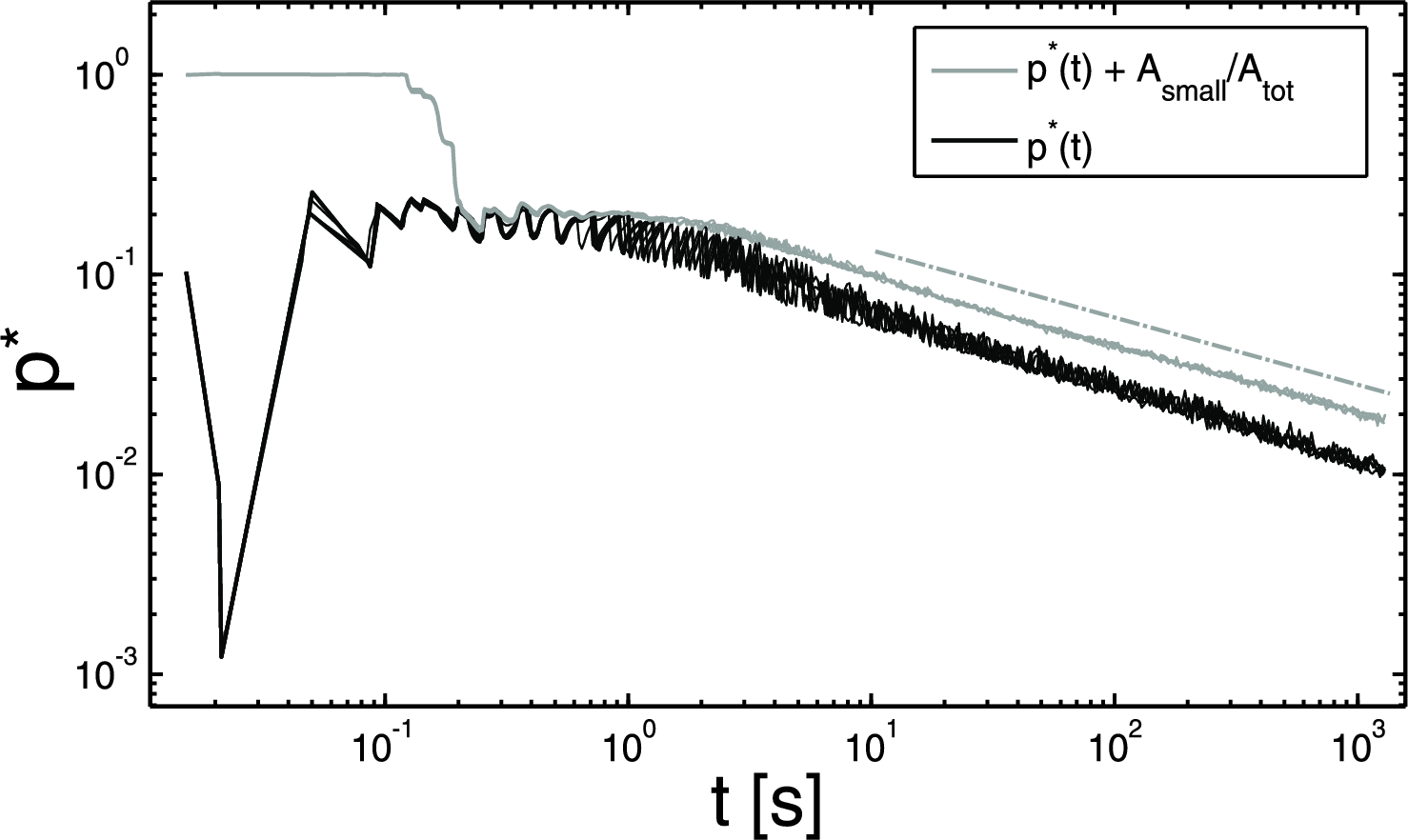}
\caption{Time evolution of the effective porosity (same parameters
as in \Fig{n density and cutoffs}). The lower lines (black)
represent the total effective porosity, as defined in \Eq{modified
porosity def}, for five simulations with different but equivalent
initial conditions; the upper lines (gray, solid) represent the
partial effective porosity $\bar{p}^* = 1 - A^*/A_{\text{tot}}$, where
$A^*$ is the sum of the areas (lengths) covered by the droplets of
size $s
> \lambda^{*3}$. The gray dashed line represent the line used to fit
the porosity $\bar{p}^*$ and it has a slope $-k = -0.332 \pm
0.003$, on the log-log axes. From this we estimate $\tau = 1.112
\pm 0.002$.}
 \label{Fig:modifiedPorosity_p*(t)}
\end{figure}

In order to improve the statistics, we run five simulations for
each case, with different but equivalent initial conditions. To
this aim, we take different seeds in the random number generator
used to create the initial distribution of the radii. The curves
$p^*(t)$ and $N(t)$ for different random numbers seeds all lie on
top of each other, and they have the same scattering of the data
(see Figs.~\ref{Fig:modifiedPorosity_p*(t)} and
\ref{Fig:N_droplets(t)}). The five runs produced data at time
instants which were not perfectly in-phase, due to the
event-driven nature of the model. Therefore, we use the curves
$p^*(t)$ and $N(t)$, produced by overlapping the individual curves
of the five runs, to extract the exponents $k$ and $z'$
respectively.

In \Fig{modifiedPorosity_p*(t)} we show the time evolution of the
modified porosity $p^*(t)$, for five equivalent initial
conditions. The following parameters were used: $\varepsilon = 2\%
$, $R_f = R_{\text{min}} = 1 \,\mu$m, $\Phi = 1000 \,\mu$m$^2
/ $s. The lower lines (black) represent the total effective
porosity, as defined in \Eq{modified porosity def}. The upper
lines (gray, solid) represent the partial effective porosity
$\bar{p}^* = 1 - A^*/A_{\text{tot}}$, where $A^*$ is the sum of the areas
(lengths) covered by the droplets of middle and large size.
Specifically, we keep into account only the droplets of size $s_c
> \lambda^{*3}$ and we use these curves to find the exponent $k$.
As pointed out at the end of \Sect{Porosity}, this choice does not
affect the resulting exponent, but it reduces the fluctuations
related to the microscopic details of the nucleation of new
droplets. The gray dashed line shows the best fit of the slope
$-k$ calculated from fitting the modified porosities $\bar{p}^*$
of the five simulations. For the considered parameters, we obtain
$k = 0.332 \pm 0.003$.

In \Fig{N_droplets(t)} we show the time evolution of the number of
droplets, for the same data of \Fig{modifiedPorosity_p*(t)}. The
dark gray thin lines (blue online) represent the total number of
droplets  $N(t)$. The black lines represent the number of the
large droplets: $s/\Sigma(t)>0.75^3$; by fitting them we estimate
a decaying exponent of $0.503 \pm 0.026 \simeq 1/2$, in agreement
with the theoretical prediction for a monodisperse droplet
populations (see \Eq{exponent z' for time decay of N(t)}, for
$\tau \leq 1$). The lit-gray thin lines (green online) depict the
number of the small droplets, with $s/\Sigma(t) < x_{N0}(t)$ and
$x_{N0}(t) = \lambda^{*3}/ \Sigma(t)$. They suffer from large
oscillations, reflecting the repopulation of large areas that are
episodically released by the merging of large droplets. The gray
thick lines (red online) show the number of droplets of middle and
large size, with $s>\lambda^{*3}$, as defined above. We use these
lines to find the exponent characterizing the decay of the number
of droplets in time $z' = 0.329 \pm 0.002$. From \Eq{exponent z'
for time decay of N(t)} and \Eq{exponent k, for time decay of
effective porosity}, we expect $k=z'$, which is verified within
the estimated error; the relative error on the average value is
$\sim 0.4\%$. From \Eq{exponent k, for time decay of effective
porosity}, we derive the polydispersity exponent of the droplet
size distribution $\tau = 1.112 \pm 0.002$, which is within the
physically acceptable range $1< \tau < \theta$.

\subsection{Droplet number density}

We calculate $n(s,t)$ from our numerical data by dividing the
range of the sizes $s$ into $N_{\text{bins}}$ bins of width
$\Delta s_j$, centered around $s_j$, where $j =
1,..,N_{\text{bins}}$. At each time instant, we then calculate
the droplet number density $n(s_j,t)$ by counting how many
droplets lie in the respective bins, $(s_j - \Delta s_j/2) < s^* <
(s_j + \Delta s_j/2)$, and dividing the respective numbers by
$(L_f \Delta s_j)$, where $L_f$ is the total length of the fiber.
We then take the average of the resulting droplet size
distributions over 10 time instants and over five simulations with
different but equivalent initial conditions, each one with initial
droplet number $N_0 = 1.5 \times 10^5$. The considered time
instants are in the self-similar regime ($t>1000$ s) and they have
been chosen in such a way to have 100 points per time decade. In
particular they belong to the time range $1000$ s $<t<2000$ s.

\begin{figure}
\centering
\includegraphics[width=0.45\textwidth]{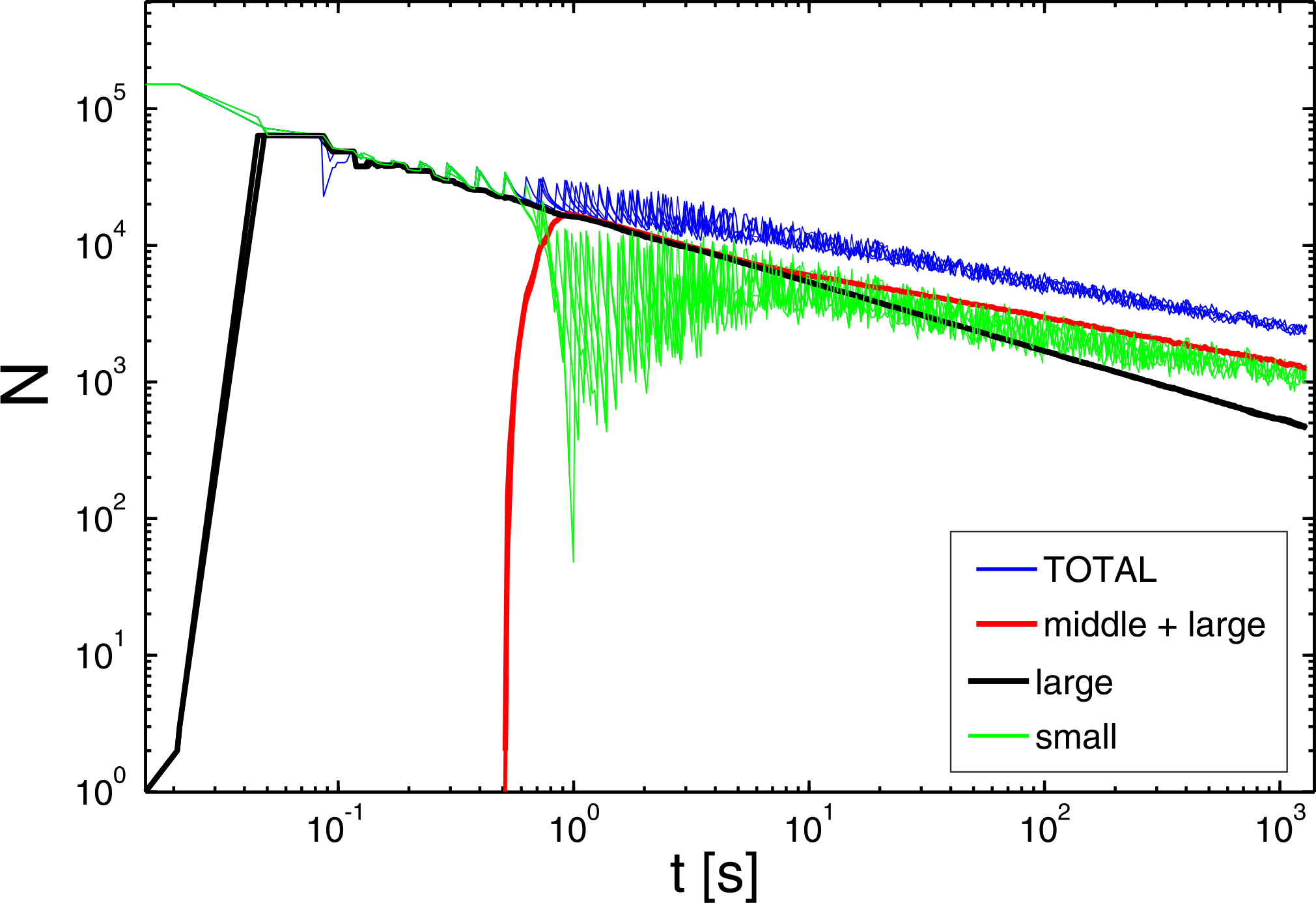}
\caption{(Color online) Time evolution of the number of droplets
per unit length  (same parameters as in \Fig{n density and
cutoffs}). The dark gray thin lines (blue online, uppermost right,
denoted as "TOTAL" in the legend) represent the total number of
droplets $N(t)$, for 5 simulations with different but equivalent
initial conditions. The lit-gray thin lines (green online)
represent the number of the small droplets, with $s \leq
\lambda^{*3}$ and its large oscillations. The mid-gray thick lines
(red online) represent the number of large and middle size
droplets, with $s > \lambda^{*3}$; upon fitting, we derive a
decaying exponent $z' = 0.329 \pm 0.002$. The black lines
represent the number of the large droplets, with
$s/\Sigma(t)>0.75^3$; their decaying exponent is $0.503 \pm 0.026
\simeq 1/2$, in agreement with the theoretical prediction for a
monodisperse droplet population.}
 \label{Fig:N_droplets(t)}
\end{figure}

The resulting distribution $n(s,t)$ is shown in \Fig{n density and
cutoffs} (black solid line). The gray solid line and the gray
dashed line represent our numerical estimate of the cutoffs for
the large droplets $\hat f(s/\Sigma(t))$ and for the small
droplets $\hat g(s/s_0)$, respectively. They are obtained from
\Eq{n(t,s)}, with the specific choice for the cutoff asymptotes
$\hat f(s/\Sigma(t)) = \hat f_0 =$const for $s/\Sigma(t) \leq
1\%$, and $\hat g(s/s_0) = 1$ for $s \geq \lambda^*$.

\begin{figure}
\centering
\includegraphics[width=0.45\textwidth]{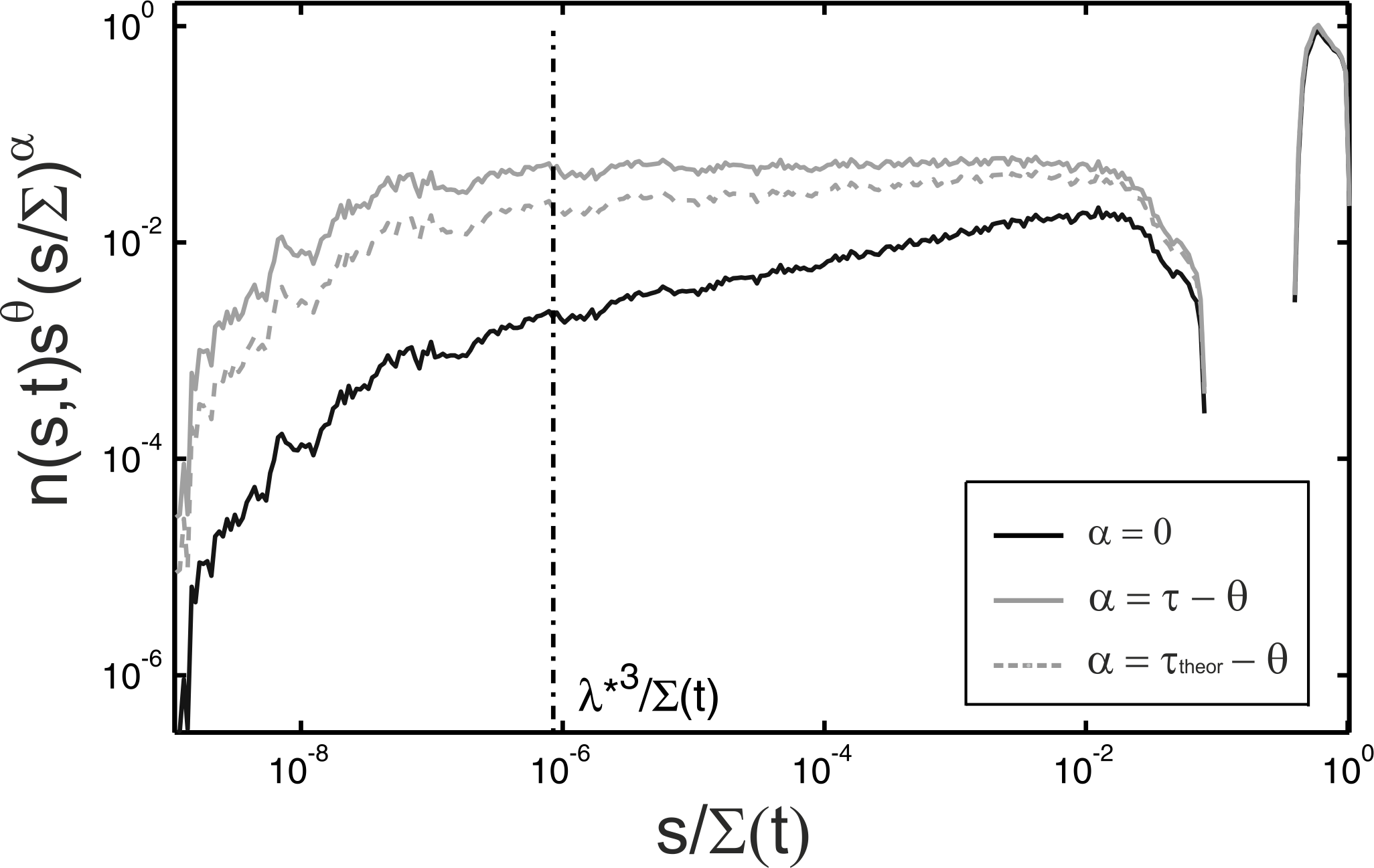}
\caption{Droplet number density $n(s,t)$ (same parameters as in
\Fig{n density and cutoffs}). The black line represents $n(s,t)
s^\theta$. The gray solid line represents the same quantity
rescaled: $n(s,t) s^\theta (s/\Sigma)^{\tau - \theta}$, with $\tau
= 1.112$, from the porosity fit: such an estimate seems to be
consistent, since the plotted curve is horizontal, in the
self-similar range (middle size droplets). The gray dashed line
represents the same rescaled quantity with $\tau_{\text{theor}}
= 7/6$ \cite{bla00}: such a value is not consistent, since the
plotted curve is not horizontal, in the range of self-similar
sizes. The present graphs are derived by averaging the droplet
size distributions over 10 time instants and over five simulations
with different but equivalent initial conditions. The considered
instants are in the self-similar regime ($t > 1000 s$) and they
are chosen in such a way to have 100 points per time decade.}
 \label{Fig:DropletNumberDensity_n(s,t)}
\end{figure}

\begin{figure*}
\[
\raisebox{0.25\textwidth}{\text{(a) \ }}
\includegraphics[height=0.27\textwidth]{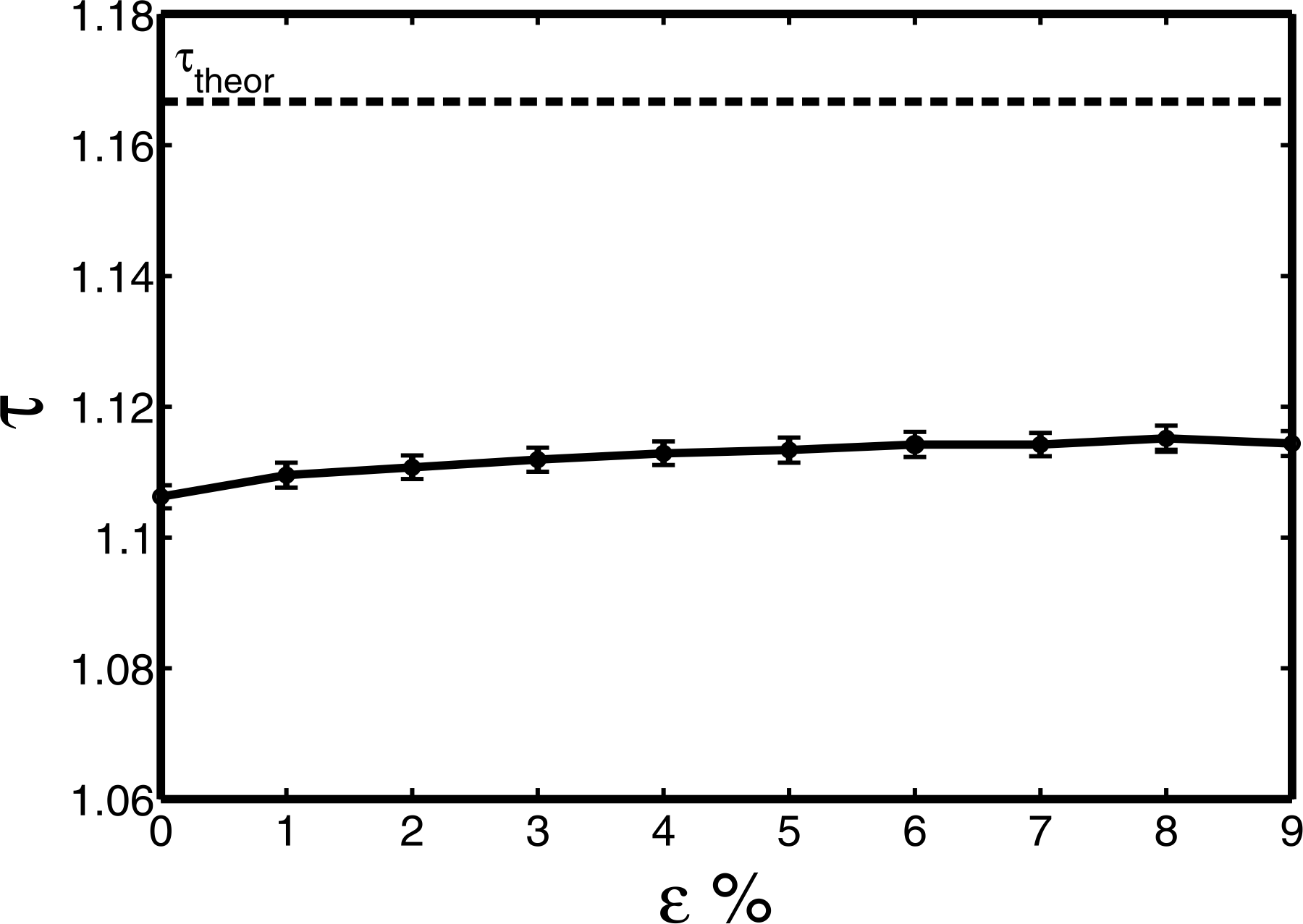}
\qquad\qquad \raisebox{0.25\textwidth}{\text{(b) \ }}
\includegraphics[height=0.27\textwidth]{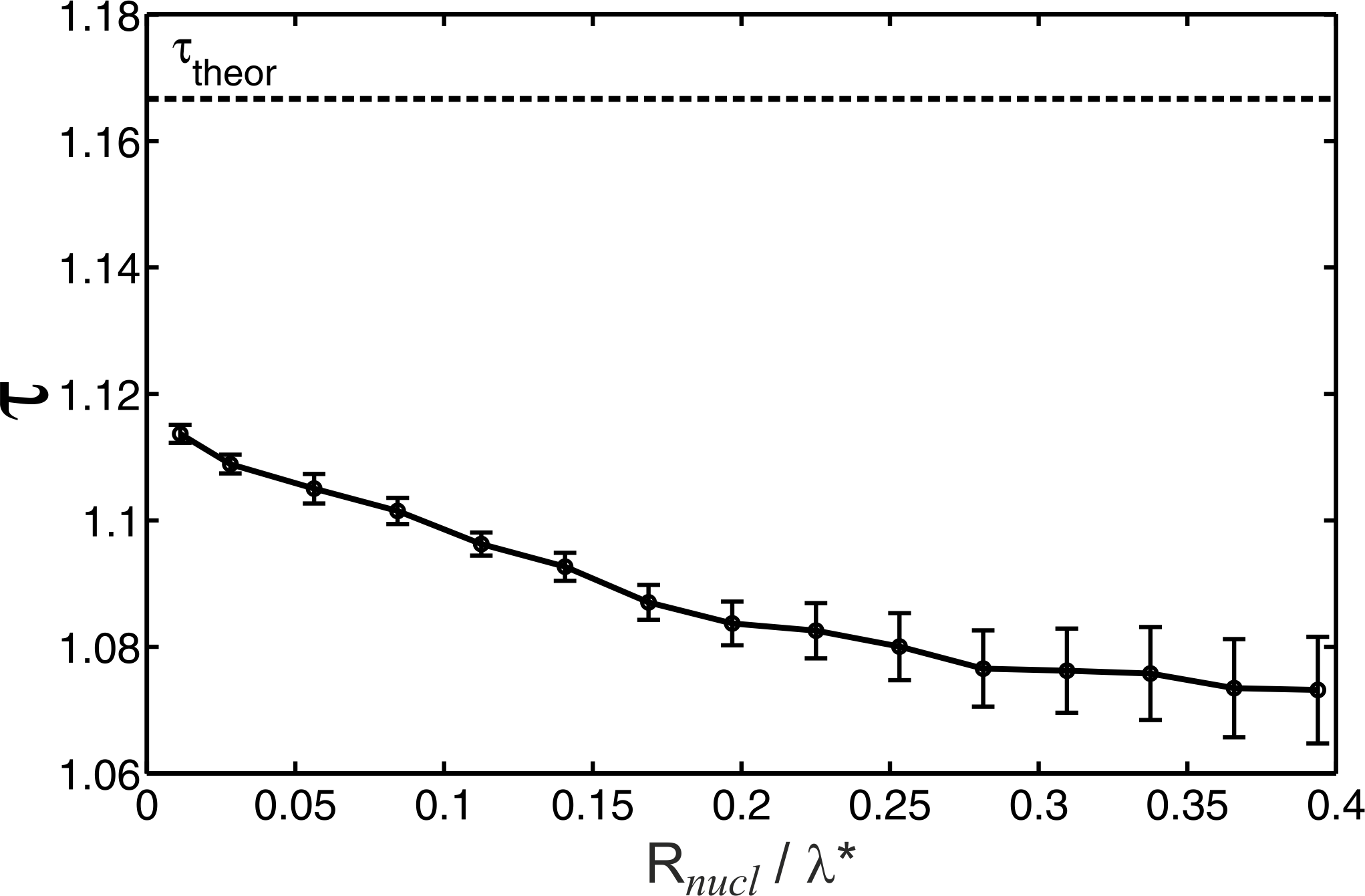}
\]
\caption{The exponent $\tau$ as a function of  (a) the interaction
range, $\varepsilon$, and (b) the ratio among the radius of the
nucleated droplets $R_{\text{nucl}}$ and the characteristic length of the
perturbation originating the droplets chain $\lambda^*$. The solid
lines represent the values of $\tau$ inferred from the fit of the
effective porosity. The dashed line is the theoretical value
predicted by the classical theory for breath figures \cite{bla00}
$\tau_{\text{theor}} = 7/6$. The parameters in (a) are
 $R_f = R_{\text{min}} = 1 \,\mu$m,
 $\Phi = 1000 \,\mu$m$^2 / $s, and
 $N_0 = 1.5 \times 10^5$,
while for (b) we used $\varepsilon = 0\%$,
 $\Phi = 1000 \,\mu$m$^2 / $s, and
 $N_0 = 1.5 \times 10^5$.
}
 \label{Fig:tau-eps-R}
\end{figure*}

In \Fig{DropletNumberDensity_n(s,t)}, we show the droplet number
density rescaled in such a way to make it dimensionless. The black
solid line represents the quantity $n(s,t) s^\theta$. With the
gray solid line we introduce a further rescaling $n(s,t) s^\theta
(s/\Sigma(t))^{\tau - \theta}$, in order to verify the estimate of
$\tau = 1.112$ from the fit of the effective porosity. The gray
dashed line depicts the same quantity, rescaled with the
theoretical prediction $\tau_{\text{theor}} = 7/6$ \cite{bla00}.
The gray solid line is horizontal, in the range of sizes
corresponding to the self-similar regime, while the gray dashed
line has a slight positive slope. Consistently, our data support a
polydispersity exponent considerably smaller than expected, but
still matching the values obtained from the decay of the porosity
\Fig{modifiedPorosity_p*(t)} and the droplet number
\Fig{N_droplets(t)}. The good agreement fully supports the
classical scaling results relating the power-law dependencies of
these quantities \cite{bey91, kol89, mea92, bla00}.

\subsection{Parameter dependence of $\tau$}

We repeated the procedure to calculate $\tau$ and to verify its
accuracy, varying several parameters, in order to clarify its
nature in terms of universality. In particular, we considered its
dependence on the interaction range $\varepsilon$ between
droplets, the nucleation radius $R_{\text{nucl}}$, the fiber
radius $R_f$, the impinging water flux per unit length $\Phi$ and
the monodispersity of the nucleated droplets chain. The variation
of the latter, has been realized by randomly redistributing
different percentages of the water volumes $V_{\text{gap},i}$
accumulated on the $i^{th}$ gap, among the new nucleated droplets
on the same gap. We found that changes in both the impinging flux
$\Phi$ and the monodispersity of the nucleated droplets do not
affect the exponent $\tau$, in line with expectations (see
\Sect{Theory}). However, the exponent $\tau$ exhibits a clear
dependence on the interaction range, as it grows with
$\varepsilon$ (see \Fig{tau-eps-R}(a), solid line). The ratio
between the nucleation radius $R_{\text{nucl}}$ and the
characteristic length of the perturbation $\lambda^*$ also has an
influence on $\tau$. In particular, $\tau$ decreases with
increasing $R_{\text{nucl}}/\lambda^*$ (see \Fig{tau-eps-R}(b),
solid line).

Remarkably, all measured values of $\tau$ are substantially
different ($4$\%--$8$\%) from the theoretical prediction \cite{bla00}
demanding $\tau_{\text{theor}} = 7/6$ (see \Fig{tau-eps-R},
dashed lines). The difference is significant, since it exceeds 10
times the estimated error of the exponent.

The dependence of the $\tau$ exponent on the interaction range
$\varepsilon$ can be explained as follows. Having larger
interaction ranges $\varepsilon$ implies having a larger area
released upon merging. One can see this by considering two merging
droplets of radius $R_1$ and $R_2$, with interaction range
$\varepsilon R_1$ and $\varepsilon R_2$ respectively. Just before
the merging, they cover an area $A = 2(R_1 + R_2)(1+\varepsilon)$.
After the merging the covered area will be $A' = 2(1+\varepsilon)
R'$, where $R'$ is the radius of the new generated droplet.
Therefore, the released area, i.e. the generated gap, will be
$\Delta g = 2(1 + \varepsilon)(R_1 + R_2 - R')$. Larger generated
gaps imply an enhancement in the nucleation process, because more
space and more water volume are available for the new droplets. On
the other hand larger gaps imply a delay in further merging
processes. Such effects are particularly relevant when the merging
droplets are large, i.e. when $(R_1 + R_2 - R')$ is large. For
increasing values of $\varepsilon$, one can then expect an
increase in the number of small droplets and a decrease in the
number of large droplets. Since $-\tau$ represents the slope of
the size distribution of the droplets $n(s,t)$ (see \Fig{n density
and cutoffs}), this will result into higher values of $\tau$ (see
\Fig{tau-eps-R}(a), solid line).

The dependence of $\tau$ on the ratio $R_{\text{nucl}}/\lambda^*$ can be
explained through the following qualitative argument. We consider
the case where the nucleation radius $R_{\text{nucl}}$ changes, but the
characteristic length of the perturbation $\lambda^*$ remains
constant, as well as all the other parameters. Given a certain
newly generated gap of size $l_i^0$, the number of possible
nucleation sites $N_i$ inside such a gap does not change with
$R_{\text{nucl}}$, but only with $\lambda^*$, since $N_i = \lfloor l_i^0
/ \lambda^* \rfloor$. The time interval $\Delta t$ during which
$N_i$ droplets can nucleate is the time required to reduce the gap
size from $l_i^0$ to $N_i \lambda^*$. After that time, only $N_i
-1$ droplets will have the space to nucleate inside the gap.
However, the water volume required to have a nucleating chain of
$N_i$ droplet is larger for larger $R_{\text{nucl}}$ and criterion
(\ref{Eq: cdt nucleation, enough vol}) may not be matched during
the time interval $\Delta t$. Therefore, for larger $R_{\text{nucl}}$, a
smaller number of new droplets will appear, before the gap closes.
Overall, a lower number of small droplets will be present in the
system, hence a larger number of big droplets, due to mass
conservation. Therefore, for increasing $R_{\text{nucl}}/\lambda^*$, the
exponent $\tau$ will decrease
(see \Fig{tau-eps-R}(b), solid line).

\subsection{Origins of the discrepancy with $\tau_{\text{theor}} = 7/6$}

\begin{figure*}
\centering
\includegraphics[width=1.0\textwidth]{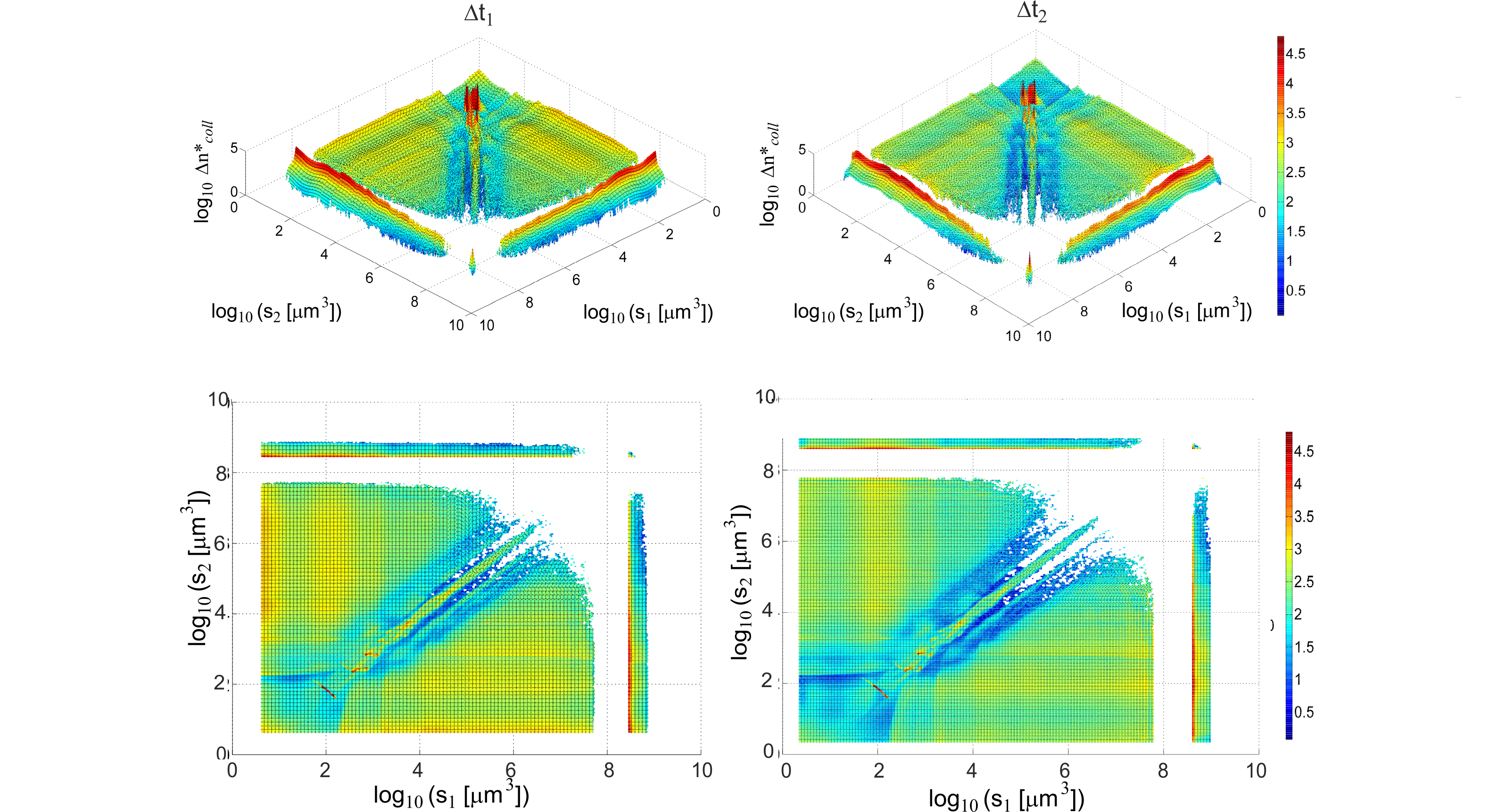}
\caption{Reduced number density of collisions per unit length
$\Delta n^*_{\text{coll}}$, during two different time intervals $\Delta
t_1 = t_1 - t_0$ (left) and $\Delta t_2 = t_2 - t_1$ (right), with
$t_0 = 480.5$ s, $t_1 = 1035.3$ s, $t_2 = 1308.9$ s (same
parameters as in \Fig{n density and cutoffs}). To a good
approximation, the plotted quantities are independent of time.
However, they show a non-trivial dependence on $s_1$ and $s_2$.}
 \label{Fig:N_coll_rescaled_in_t1-t2}
\end{figure*}


The theoretical prediction \cite{bla00}, $\tau_{\text{theor}} =
7/6$, is based on the assumption that the collision rate between
two droplets of size $s_1$ and $s_2$ can be factorized as
described in \Eq{Probability of collision, factorization}. When
one adopts the hypothesis that $\tau$ takes a universal value,
this approximation can be conveniently chosen to calculate the
values of $\tau$. When $\tau$ depends on the microscopic details
of the dynamics, as observed in \Fig{tau-eps-R}, such an
assumption must be tested for the data at hand.

To determine the collision rate for our numerical data, we count
the number of collisions per unit length per unit size in two
different time intervals $\Delta t_1 = t_1 - t_0$ and $\Delta t_2
= t_2 - t_1$, both in the self-similar regime of the droplets. In
order to have a better statistics, we average the sets of results
from five simulations with equivalent initial conditions. These
numerical data are then compared to the integral of
\Eq{Probability of collision, factorization} between $t$ and $t +
\Delta t$. Upon substitution of the expressions for $n(s,t)$ and
$N(t)$ given by \Eqs{n(t,s)} and \ref{Eq: N(t), total}
respectively, we find
\begin{widetext}
\begin{subequations}
\begin{eqnarray}
  \Delta n_{\text{coll}}(s1,s2; t, \Delta t)
  &\sim&
  \tilde{H} \; (s_1 s_2)^{(1-3\tau)/D} \; (s_1^{1/D} + s_2^{1/D})
  \left[(t + \Delta t) ^{3\tau/2 - 1} - t^{3\tau/2 - 1} \right]
  \\[2mm]
  \text{where }
  \tilde{H}
  &=&
  \hat{f} \left(\frac{s_1}{\Sigma(t)}\right) \;
  \hat{g}\left(\frac{s_1}{s_0}\right) \;
  \hat{f}\left(\frac{s_2}{\Sigma(t)}\right) \;
  \hat{g}\left(\frac{s_2}{s_0}\right) \,.
\end{eqnarray}
\label{Eq: factorization N collisions}%
\end{subequations}%
Consequently, \Eq{Probability of collision, factorization} implies
that
\begin{equation}
\Delta n_{\text{coll}}^* = \frac{\Delta n_{\text{coll}}(s1,s2;
t, \Delta t)}
 { \tilde{H} \; (s_1 s_2)^{(1-3\tau)/D} \; (s_1^{1/D} + s_2^{1/D}) \left[(t + \Delta t) ^{\frac{3\tau}{2} - 1} - t^{\frac{3\tau}{2} - 1}
 \right] }
  = \text{const} \, .
\label{Eq: reduced collision rate}
\end{equation}%
\end{widetext}%
Hence, the reduced number density of collisions per unit length of
substrate $\Delta n_{\text{coll}}^*$ should be independent of the
time $t$ and the sizes of the colliding droplets $s_1$, $s_2$. In
\Fig{N_coll_rescaled_in_t1-t2} we plot the reduced number density
of collisions per unit length of substrate $\Delta
n_{\text{coll}}^*$ during the intervals $\Delta t_1$, $\Delta
t_2$. In particular, we use the same set of data of \Fig{n density
and cutoffs} and we take $t_0 = 480.5$ s, $t_1 = 1035.3$ s , $t_2
= 1308.9$ s. We find that $\Delta n_{\text{coll}}^*$ is indeed
invariant in time, to a good approximation (see \Fig{Ncoll
rescaled, cross sections}). However, in variance with the assumed
expression for the collision rate, \Eq{Probability of collision,
factorization}, it depends on $s_1$ and $s_2$. In \Fig{Ncoll
rescaled, cross sections}(a) we show a cross section of
\Fig{N_coll_rescaled_in_t1-t2} along the main diagonal $s_1 = s_2
= s$, both for the first time interval $\Delta t_1$ (dashed line)
and the second time interval $\Delta t_2$ (solid line). In other
words, we display the reduced number of collisions between
droplets of the same size. The data present pronounced, very sharp
peaks, which do not move in time. The first peak, at $s^*_0$,
corresponds to an enhanced probability of having a collision
between droplets originating at the same moment, upon nucleation
in the same gap, when they reach the size $s^*_0=(\lambda^*/2)^3$
(see \Fig{Sketch Rmax}). The second highest peak corresponds to an
enhanced probability of collision between droplets of size
$s^*_1=\lambda^{*3}$. Following the mechanism described in
\Fig{Sketch Rmax}, we infer that this peak accounts for collisions
between droplets generated at the same moment in the same droplet
chain, that have already collided once. An interpretation along
the same lines can be given for the peaks appearing at
$s^*_2=(2\lambda^*)^3$ and $s^*_3=(4\lambda^*)^3$. With our
simulation parameters, $s^*_0$ = 87 $\mu$m$^3$ and $s^*_1$ = 700
$\mu$m$^3$, $s^*_2$ = 5613 $\mu$m$^3$ and $s^*_3$ = 44900
$\mu$m$^3$. The intermediate peaks appearing between the main ones
can be related to alternative merging mechanisms between droplets
originated at the same moment in the same droplet chain. For
example, the peak between $s^*_0$ and $s^*_1$, corresponds to a
size $s^*_{1/2}\simeq(3/2 \lambda^*)^3$, suggesting a mechanism
where every second droplet of the chain has already merged with
its neighbor and the next merging takes place between droplets of
alternated sizes $\lambda^*/2$ and $\lambda^*$. For colliding
droplets larger than $s^*_3$, such mechanisms are not relevant
anymore, thus suggesting the appearance of a self-similar area,
where the system has lost memory of the microscopic details of the
nucleation.

\begin{figure*}
\centering
\includegraphics[width=0.5\textwidth]{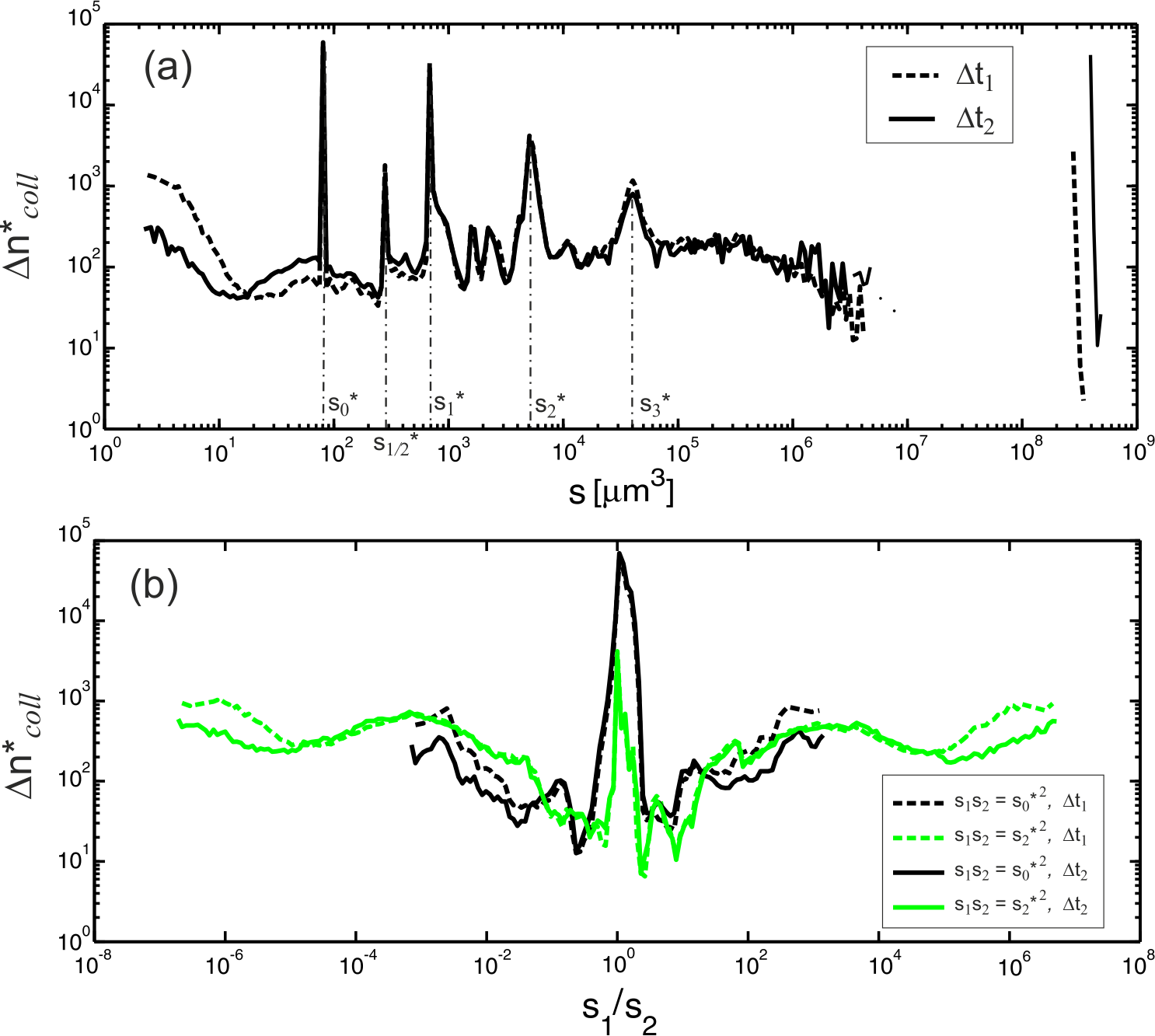}
\caption{(Color online) Cross sections of the data displayed in
\Fig{N_coll_rescaled_in_t1-t2} (a) along the main diagonal and (b)
perpendicularly to it, during two different time intervals $\Delta
t_1$ (dashed lines) and $\Delta t_2$ (solid lines). (a) Along the
main diagonal, sharply defined maxima appear at sizes $s_i^*$
(with $i = 0,\frac{1}{2},1,2,3$). (b) Perpendicularly to the main
diagonal, along the lines $ s_1 s_2= (s_0^*)^2$ (black lines) and
$s_1 s_2= (s_2^*)^2$ (lit-gray lines, green online), there are
maxima at the diagonal $s_1/s_2 = 1$, dips to its left and right
and a plateau for $|\log_{10} s_1/s_2| \gtrsim 3$.}
 \label{Fig:Ncoll rescaled, cross sections}
\end{figure*}

\Fig{Ncoll rescaled, cross sections}(b) displays further cross
sections of \Fig{N_coll_rescaled_in_t1-t2} during $\Delta t_1$
(dashed lines) and $\Delta t_2$(solid lines). These cross sections
are perpendicular to the main diagonal, and they show the peaks
appearing \Fig{Ncoll rescaled, cross sections}(a) at $s^*_0$
(black lines) and $s^*_2$ (lit-gray lines, green online). The data
display a recurring structure, with a peak corresponding to the
main diagonal, and a dip just next to it. Such a structure is
present along the whole main diagonal, as soon as the droplets are
large enough to allow for secondary nucleation in the emerging
gaps. The peak corresponds to the collisions between droplets of
the same size. The dip surrounding the peak can be explained
through the concept of self-similarity itself. The global droplet
size distribution $n(s,t)$ can be regarded as a superposition of
bimodal droplet size distributions, with the same shape displayed
in \Fig{n density and cutoffs}, but different values of $\Sigma$.
If we consider a generic portion of the area occupied by the
droplets, where the largest size is $\Sigma^*$, the rescaled
droplet size distribution of such an area will present several
droplets of similar size $\Sigma^*$ (see the bump in
\Fig{DropletNumberDensity_n(s,t)}), from which, the peaks centered
on the main diagonal in \Fig{N_coll_rescaled_in_t1-t2} originate.
Such local droplet size distribution will also present a gap
similar to the one of \Fig{n density and cutoffs}, from which the
dips of \Fig{Ncoll rescaled, cross sections}(b)
 originate.

We conclude that the factorization of the collision rate
$\dot{n}_{\text{coll}}$ \Eq{Probability of collision,
factorization}, based on the assumption of the distribution of the
droplets sizes $s_1$ being uncorrelated to the distribution of the
sizes of the neighbors $s_2$, does not hold for our data.

\section{Conclusions}
\label{sec:Conclusion}

So far, only few numerical \cite{der90, der91, mea91} and
experimental works \cite{ste90} have specifically addressed breath
figures on a one-dimensional substrate. Performing repeatable and
controllable experiments for droplets on a fiber presents
technical difficulties, such as keeping the temperature of a thin
fiber constant. Quasi-one-dimensional settings have been realized
by means of scratches on a plate \cite{ste90}, which allowed for a
better control of the temperature of the substrate.

From the classical theory of breath figures it is known that the
size distribution of droplets on a substrate becomes self-similar
after some time that the system evolves; therefore it can be
described by means of scaling laws, at least in the intermediate
range of droplets sizes (the polydisperse range). In particular,
two scaling exponents appear: $\theta$ and $\tau$. The value of
$\theta$ can be inferred from dimensional analysis, while the
value of the polydispersity exponent $\tau$ is non-trivial. The
present work investigated the dependence of this exponent on the
microscopic details of the system, in order to verify the recent
finding \cite{bla12} that $\tau$ might not be universal, as
commonly assumed. We developed an event-driven model for
three-dimensional droplets on a one-dimensional substrate (fiber).
We included the following details: the growth of the droplets by
mass deposition from an impinging flow of supersaturated vapor,
the precursor film between adjacent droplets, the nucleation
process by surface tension driven instability and the merging of
adjacent droplets. We calculated $\tau$, as well as the exponents
characterizing the decay in time of the porosity and the number of
droplets. The relations among these three exponents derived in the
classical scaling analysis \cite{bey91, kol89, mea92, bla00}, hold
also for our model. As expected, we found that the exponent $\tau$
does not depend on the impinging mass flow and on the
monodispersity level of the nucleation process. However, it does
depend on the interaction range $\varepsilon$ of the droplets and
on the ratio between the nucleation radius $R_{\text{nucl}}$ and
the spacing $\lambda^*$ between the nucleating droplets. In
particular, $\tau$ grows with increasing interaction ranges
$\varepsilon$ and decreasing ratios $R_{\text{nucl}}/ \lambda^*$.
Our results contradicts the expectation that the exponent $\tau$
should be universal. Additionally, the values of $\tau$ that we
inferred from our simulations differ by 10 standard deviations
from the theoretical prediction $\tau_{\text{theor}} = 7/6$
\cite{bla00}. Such a prediction was based on the assumption that
the probabilities of having two neighboring droplets of prescribed
sizes are uncorrelated. We analyzed the distribution of the
collisions respect to the sizes of the colliding droplets and we
showed that this assumption is inaccurate when one keeps into
account the microscopic details of the system.

Our observations pose new questions on the non-universal nature of
the polydispersity exponent $\tau$, such as the specific
mechanisms underlying its parametric dependence as well as its
quantitative values. These questions have been raised here for a
setting where the flux impinging on the droplets is proportional
to the wetted length on the fiber, i.e. a situation where the
droplet growth is limited by the diffusive mass transport to the
fiber. Alternatively, one could also consider a setting where the
droplets grow in proportion to their total surface area. The
classical expectation was that this change of the microscopic
details of the droplet growth should not affect the value of
$\tau$. In order to test this expectation and to address the
emerging questions, further numerical simulations should be
developed, as well as a new theoretical framework accounting for
the non-universal nature of $\tau$.

\begin{acknowledgments}

  We acknowledge discussions with Johannes Blaschke, Philipp
  D\"onges, Gunnar Kl\"os, Artur Wachtel, and Tobias Lapp, as well as feedback on the
  manuscript by Stephan Herminghaus.

\end{acknowledgments}



%

\end{document}